\documentclass[symmetry,review,accept,moreauthors,pdftex]{Definitions/mdpi} 

\firstpage{1} 
\pubvolume{1}
\issuenum{1}
\articlenumber{0}
\pubyear{2021}
\copyrightyear{2021}
\externaleditor{{Academic Editor(s)}: Angel Gómez Nicola and Jacobo Ruiz de Elvira} 
\datereceived{22 September 2021} 
\dateaccepted{26 October 2021} 
\datepublished{} 
\hreflink{https://doi.org/} 
\pdfoutput=1

\usepackage[british]{babel}
\usepackage{amsmath,amssymb}
\usepackage{amsfonts}
\usepackage[matha,mathb,mathx]{mathabx}
\usepackage{cancel}
\usepackage{epsfig}
\usepackage{epstopdf}
\usepackage{graphicx,psfrag}
\usepackage{subcaption}
\usepackage{enumitem}

\newcommand{\beq}{\begin{equation}}
\newcommand{\eeq}{\end{equation}}
\newcommand{\bea}{\begin{eqnarray}}
\newcommand{\eea}{\end{eqnarray}}
\newcommand{\nn}{\nonumber}
\newcommand{\eq}{Equation~}

\newcommand{\fig}{Figure~}

\newcommand{\bx}{{\bf x}}

\def\lsi{\raise0.3ex\hbox{$<$\kern-0.75em\raise-1.1ex\hbox{$\sim$}}}
\def\gsi{\raise0.3ex\hbox{$>$\kern-0.75em\raise-1.1ex\hbox{$\sim$}}}

\newcommand{\order}[2][]{\mathcal{O}#1(#2#1)}

\newcommand{\coeff}[2]{\mathcal{\uppercase{#1}}_{#2}}

\newcommand{\NTau}{N_\tau}

\newcommand{\Nf}{N_\text{f}}

\newcommand{\psibarpsi}{\bar{\psi}\psi}
\newcommand{\chiralcond}{\langle \psibarpsi \rangle}

\newcommand{\mud}{m_{u,d}}
\newcommand{\Action}{\mathcal S}

\newcommand{\Temp}{\ensuremath{T}}
\newcommand{\Tc}{\ensuremath{\Temp_c}}

\Title{Lattice Constraints on the QCD Chiral Phase Transition at Finite Temperature and Baryon Density}

\TitleCitation{Lattice Constraints on the QCD Chiral Phase Transition at Finite Temperature and Baryon Density}

\Author{Owe Philipsen 
 $^{\orcidA{}}$ }

\AuthorNames{Owe Philipsen}

\AuthorCitation{Philipsen, O.}

\address[1]{%
ITP, Goethe-Universit\"at Frankfurt am Main, Max-von-Laue-Str. 1, 60438 Frankfurt am Main, Germany; philipsen@itp.uni-frankfurt.de}

\abstract{The thermal restoration of chiral symmetry in QCD is known to proceed by an analytic crossover, which
is widely expected to turn into a phase transition with a critical endpoint as the baryon density is increased. 
In the absence of a genuine solution to the sign problem of lattice QCD, simulations at zero and imaginary 
baryon chemical potential in a parameter space enlarged by a variable number of quark flavours and quark masses 
constitute a viable way to constrain the location of a possible non-analytic phase transition and its critical endpoint.
In this article I review recent progress towards an understanding of the nature of the transition in the massless limit,
and its critical temperature at zero density. 
Combined with increasingly detailed studies of the physical 
crossover region, current data bound a possible critical point to $\mu_B$ $\gsi$ $3T$.   
}


\begin{document}
\section{Introduction}

Many salient features of the spectrum of hadrons and their interactions observed in nature
are determined by the near-chiral symmetry of the QCD Lagrangian and its spontaneous breaking by the 
QCD vacuum.
The fact that nature breaks chiral symmetry also explicitly by non-vanishing quark masses can be
taken into account in terms 
of chiral perturbation theory~\cite{Gasser:1983yg,Leutwyler:1994fi}, 
due to the smallness of the $u$- and $d$-quark~masses.  

In a hot and/or dense medium, chiral symmetry is expected to be gradually restored once the temperature
exceeds $T$ $\gsi$ $160$ MeV or the baryon chemical potential \mbox{$\mu_B$ $\gsi$ $1$ GeV}. Associated with this
change of the realised symmetry, one also expects a change of dynamics and its underlying degrees of freedom which,
asymptotically, should become quarks and gluons. Similar to the vacuum properties of the theory, 
one also expects the properties of this thermal transition to be closely related to those of the 
theory in the massless~limit. 

The low energy scales of the transition region demand a
non-perturbative first principles approach like lattice QCD. Unfortunately, a~severe sign problem prohibits simulations 
by importance sampling for non-vanishing chemical potential, 
for introductions see~\cite{Gattringer:2010zz,Philipsen:2010gj,Aarts:2015tyj}. 
Despite tremendous efforts over several decades, no genuine solution to this problem is available to date, and~
knowledge of the QCD phase diagram by direct calculation remains scarce. Nevertheless, during~the last years
there has been considerable progress in studying the properties of the thermal QCD transition in an enlarged
parameter space, where the number of flavours and quark masses is varied 
away from their physical values. Besides~giving theoretical insight into the interplay of symmetries and dynamics
in controlled situations, such studies also provide constraints on the physical phase diagram, which are beginning to 
become phenomenologically relevant.  
In this article, I collect some recent results allowing to better understand and remove lattice artefacts, resulting 
in a consistent picture that also accommodates apparently conflicting statements based on earlier simulations. 
In particular, I will
focus on developments suggesting a modified version of the Columbia plot and constraints on the location of a 
possible critical~point. 
 
\section{Some Lattice~Essentials}

For readers less familiar with lattice QCD, it  may be useful to recall a number of basic definitions and issues
that are important for the interpretation of the following lattice results. For~details, see the corresponding 
textbooks~\cite{Gattringer:2010zz,Montvay:1994cy,DeGrand:2006zz}. 
Conventionally one considers Euclidean space-time 
discretised by hypercubic lattices with $N_s^3\times \NTau$ points separated by a lattice spacing $a$. This represents
a box with spatial length $L=N_s a$ and Euclidean time extent $a\NTau=1/T$, which corresponds to inverse temperature
as in continuum thermal field theory, once (anti-)periodic boundary conditions for (fermionic) bosonic fields
have been applied. When a quantum field theory is formulated on the lattice, its action (and all observables)
differ from those in the continuum.
As an example, consider Wilson's $SU(N_c)$ pure gauge action
\beq
S_{YM}=\beta \sum_p \Big[1-\frac{1}{N_c}{\rm Tr}U(p)\Big]\;,\quad \beta=\frac{2N_c}{g^2}\;,
\label{eq:wilson}
\eeq
which is constructed from so-called plaquette variables $U(p)$. These are the fundamental squares of the lattice consisting of four 
gauge links $U_\mu(x)=\exp(-igaA_\mu(x))$, which are group elements of $SU(N_c)$. For~sufficiently small lattice spacing
the plaquette can \mbox{be expanded},
\beq
U(p)=1+ia^2gF_{\mu\nu}-\frac{a^4g^2}{2}F_{\mu\nu}F_{\mu\nu}+O(a^6),
\eeq  
and the classical action (\ref{eq:wilson}) is seen to reproduce the continuum Yang-Mills action in the limit $a\rightarrow 0$. However: for any finite lattice spacing $a$ the action differs by terms of $O(a^2)$ from the continuum action. 
Such terms disappearing in the continuum are called ``irrelevant'', at~finite lattice spacing they constitute lattice artefacts. 
An improvement of lattice actions in the sense of reducing such artefacts~\cite{Symanzik:1983dc} can be achieved by adding 
irrelevant terms to the lattice action, which subtract those already present. Quite generally, lattice actions
are then not unique but may differ by arbitrary irrelevant terms. All such differences between valid actions and observables must disappear in
the continuum~limit.  

Approaching the continuum limit is a highly non-trivial affair, as~one has to take $a\rightarrow 0$ while keeping other
dimensionful quantities, like masses or temperature, fixed along a so-called line of constant physics in the lattice
parameter space. The~lattice spacing is tuned indirectly through the running gauge coupling evaluated at the cutoff scale,
$g=g(\Lambda\sim a^{-1})$, so that by 
asymptotic freedom $g(a\rightarrow 0)\rightarrow 0$. This implies that the lattice gauge coupling
$\beta(a\rightarrow 0)\rightarrow \infty$. For~thermodynamics, keeping temperature constant during the continuum approach
implies $\NTau\rightarrow \infty$ in the continuum limit. 
Thus, while the precise value of the lattice spacing is determined by some form of
scale setting, finer thermodynamics lattices imply larger $\NTau$, and~lattice artefacts can be parametrised in units
of temperature as powers of the dimensionless $aT=\NTau^{-1}$.

The most problematic lattice artefacts for QCD by far are those afflicting the fermion formulation. In~particular, it is impossible
to put chiral fermions on a lattice without artificial doubler degrees of freedom while maintaining locality~\cite{Nielsen:1981hk}. 
It is clear
that this implies severe caveats for any investigation of chiral symmetry breaking. In~a nut shell, the~current choices
are:
\begin{itemize}
\item The Wilson formulation: By adding irrelevant mass terms $\sim a^{-1}$, the~doubler degrees of freedom become heavy
and decouple in the continuum limit. However, for~any finite lattice spacing chiral symmetry is broken completely by these terms.
\item The staggered formulation: Spin and flavour degrees of freedom are distributed to different lattice sites, which 
can be done so as to reduce the number of doublers. 
The remaining ones are removed by taking an appropriate root of the fermion determinant,
which can only be fully valid in the continuum. In~this formulation the original chiral symmetry is reduced
from $SU(2)_L\times SU(2)_R\rightarrow O(2)$.
\item Formulations with a lattice version of full chiral symmetry exist in the form of domain wall or overlap fermions,
and are expected to eventually supersede the previous formulations.
However, they require complicated non-local constructions and currently are more expensive to simulate by over an 
order of magnitude.
\end{itemize}   

Due to the enormous numerical cost of thermodynamical investigations for light fermions, most lattice results
covered here are based on different versions of the staggered or Wilson discretisations. In~order to avoid the damage these
do to chiral symmetry and its spontaneous breaking, it is mandatory to first take the continuum limit to remove
the lattice artefacts, and~only then approach the chiral~limit.  

A general difficulty in investigating phase diagrams is the fact that non-analytic phase transitions only
exist in infinite volume~\cite{Yang:1952be,Lee:1952ig}, 
while numerical simulations are obviously limited to finite boxes.  
A change of dynamics associated with a phase transition is signalled by a 
rapid change of the expectation values of suitable observables $O(\bx)$,
accompanied by maximal fluctuations, i.e.,~peaks in their susceptibilities,
\beq
\chi_O=\int d^3x \;\Big( \langle O(\bx)O(\mathbf{0})\rangle -\langle O(\bx)\rangle \langle O(\mathbf{0})\rangle\Big)\;.
\eeq 
This allows to determine the location of a transition as a function of the bare parameters of the theory. 
In a finite box this is always an analytic crossover, and~the behaviour on different volumes has to be compared to 
see whether a non-analytic phase transition builds up in the thermodynamic limit. The~peak height will diverge at a second-order transition 
with $\sim V^\sigma$, where $\sigma$ is a combination of critical exponents appropriate for the 
observable and the universality class of the transition. For~a first order transition the divergence is $\sim V$, whereas 
for a crossover the peak height will saturate at a finite value.  
More intricate (and reliable) finite size scaling studies are based on the first three generalised cumulants ($n=2,3,4$ )
of the fluctuation of an observable, $\delta O=O-\langle O\rangle$,
\beq
B_n(\delta O)=\frac{\langle (\delta O)^n\rangle}{\langle (\delta O)^2\rangle^{n/2}}\;, 
\eeq
which besides the peak height also quantify the skewness and the tails of the statistical distribution of the observable,
thus giving access to various critical~exponents.

Finally, the~infamous sign problem at finite baryon density appears  because of a property of the fermion determinant,
defined with complex chemical potential for baryon number
(in continuum notation),
\bea
\det(\cancel{D}+m-\mu\gamma_0)&=&{\det}^*(\cancel{D}+m+\mu^*\gamma_0)\;.
\eea
Thus, the~fermion determinant is complex for real chemical potential, but~stays real for purely imaginary chemical potential.
For real chemical potential, all imaginary parts cancel out of the partition function exactly, but~the real part is oscillatory and
averages to the so-called average sign, evaluated by the partition function with $\mu=0$,
\beq
Z(\mu)=Z(0)\left\langle \frac{\det(\mu)}{\det(0)}\right\rangle_0 =Z(0) \exp \Big(-V(f(\mu)-f(0))/T\Big)\;.
\eeq
This is also the simplest reweighting factor multiplying any observable, when a $\mu=0$ ensemble is reweighted to $\mu\neq 0$.
Its size is governed by the free energy difference of the systems with and without chemical potential, and~is exponentially
suppressed for large volumes. Hence all expectation values of observables are lost in the statistical errors 
sooner or later. It must be stressed that this is only a problem for numerical 
evaluations by importance sampling, other evaluations and the physics itself have no issue with chemical 
potential. For~more details, see introductions  like~\cite{Gattringer:2010zz,Philipsen:2010gj,Aarts:2015tyj}. So far there is no genuine algorithmic cure for the sign problem, 
for an overview of earlier attempts see
~\cite{deForcrand:2009zkb} and the proceedings of the annual lattice conferences for~updates. 

\section{The Columbia Plot at Zero Baryon~Density}

\begin{figure}[H]
\includegraphics[width=0.35\textwidth]{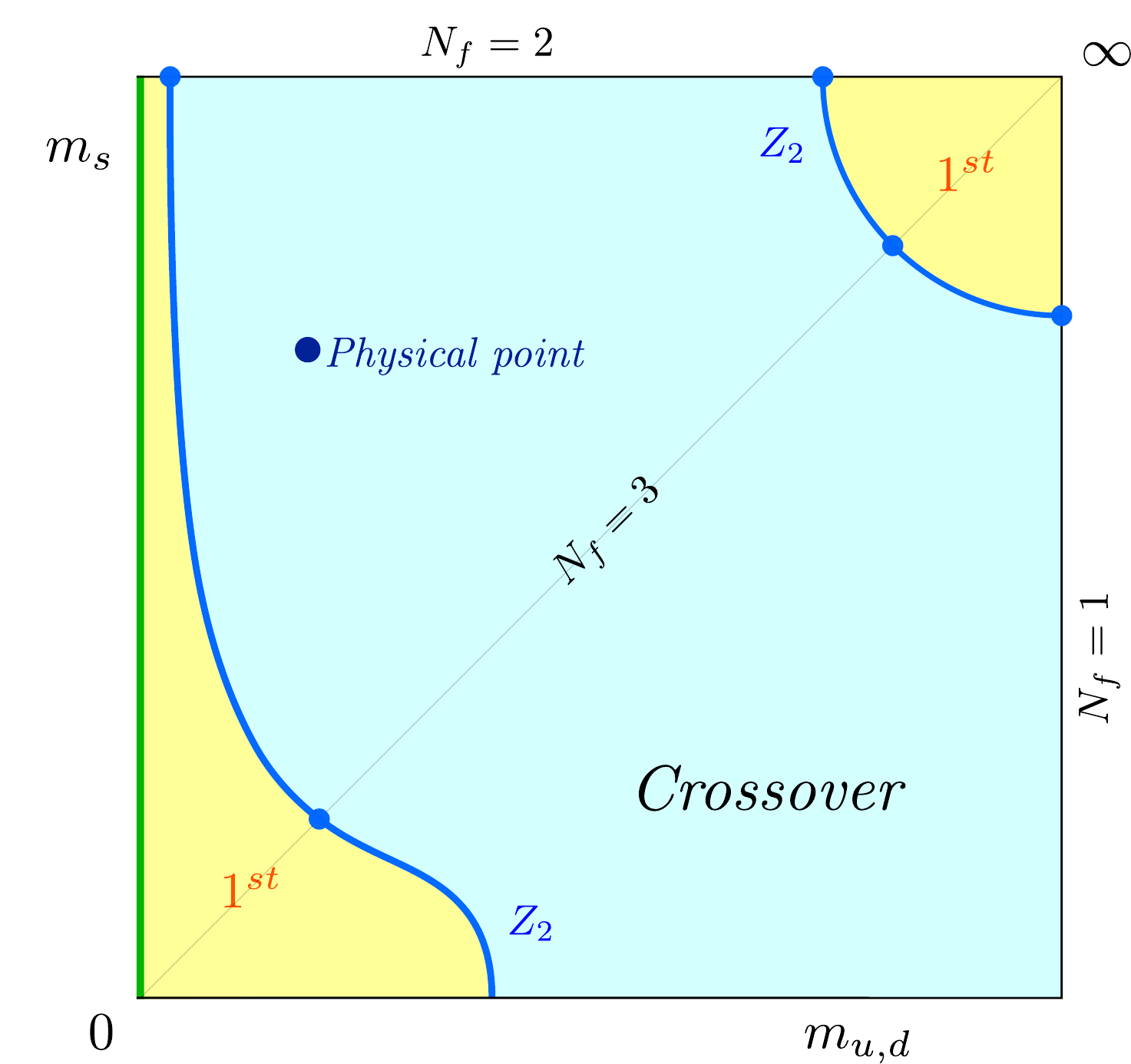}
\includegraphics[width=0.35\textwidth]{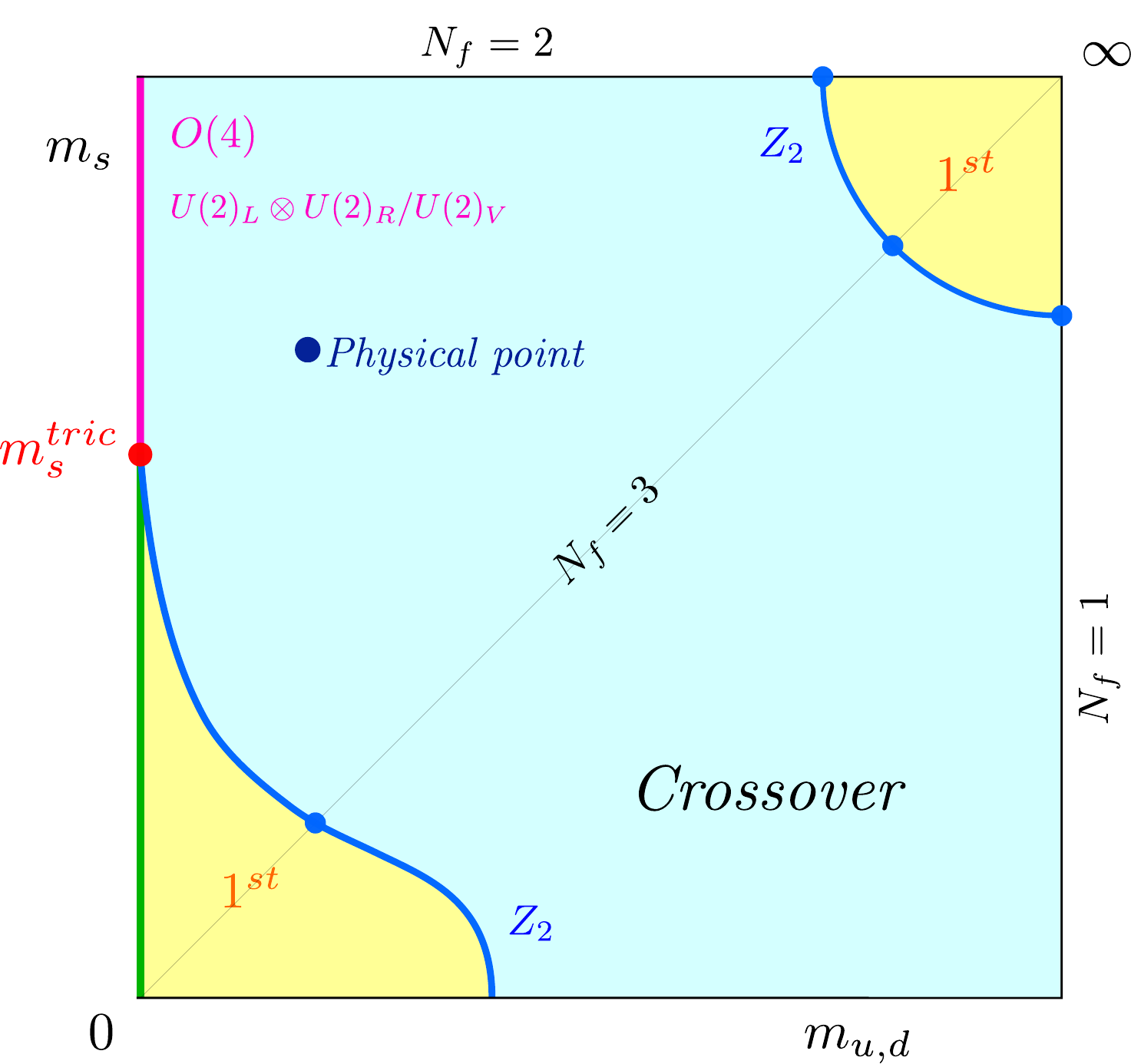}
\caption[]{Possible scenarios for the nature of the thermal QCD transition as a function of the quark masses,
with a first-order (left) or a second-order (right) transition in the chiral limit of the light quarks. Every point 
of the plot represents a phase boundary, with~an implicitly
associated (pseudo-)critical temperature $T_c(m_{u,d},m_s)$. \protect (Here and in the following, 
 ``phase boundary'' refers to a (pseudo-)critical combination of parameters
irrespective of the nature of the transition, which can be first order, second order or crossover.)} 
\label{fig:columbia}
\end{figure}

As a point of departure, consider the thermal QCD transition at zero baryon density, $\mu_B=0$.
The nature of this transition for physical quark masses has been known for some time to be an analytic 
crossover~\cite{Aoki:2006we}.
Away from the physical point,
the order of the QCD thermal
transition with $N_f=2+1$ quarks as a function of quark masses is usually specified in a so-called Columbia 
plot~\cite{Brown:1990ev}, for~which two possible realisations, to~be discussed below, are shown in \fig \ref{fig:columbia}.

\subsection{The Deconfinement~Transition}

Even though it is far from the physical point or the chiral limit, the~heavy mass corner is interesting 
in its own right and promises insights, which will also be useful for light quarks. 
In the quenched limit QCD reduces to $SU(3)$ Yang-Mills theory in
the presence of static quarks, whose propagators are Polyakov loops which represent an order parameter
for the global $Z(3)$ center symmetry related to confinement~\cite{McLerran:1981pb}.
Its spontaneous breaking at some critical temperature proceeds by a first-order deconfinement phase 
transition~\cite{Boyd:1996bx}, whose latent heat has recently also been computed~\cite{Shirogane:2016zbf}.
In the presence of dynamical quarks, the~center symmetry is explicitly broken by $1/m_q$ and the first-order 
phase transition weakens with diminishing quark masses, 
until it disappears along a second-order critical line 
in the 3d $Z(2)$ universality class. Contrary to the chiral limit, this entire region 
can be simulated directly and a complete non-perturbative understanding should be attainable in the near future.
The location of the critical line is under 
investigation~\cite{Saito:2011fs,Ejiri:2019csa,Cuteri:2020yke,Kiyohara:2021smr}. 
While continuum extrapolations are
not yet available, the~presently finest lattices predict the pseudo-scalar mass evaluated on the critical point for $N_f=2$ 
to be about $m^c_{PS} \sim 4$ GeV~\cite{Cuteri:2020yke}. Together with the latent heat, these quantities should permit
a detailed understanding of the confinement-deconfinement transition
and its interplay with screening by dynamical quarks, which 
will be valuable for the physics interpretation of the crossover region at the physical point.
Finally, these quantities 
can serve as benchmarks to test and tune truncations involved in functional renormalisation group or
Dyson-Schwinger approaches~\cite{Fischer:2014vxa}, or~effective lattice
theories for the heavy quark region, which are employed to describe 
finite density physics~\cite{Fromm:2011qi,Fromm:2012eb}.

\subsection{The Chiral Transition at Zero Baryon~Density}

The situation is far more difficult in the opposite limit of massless quarks, because~it cannot be simulated directly. 
For several decades expectations have been based on       
an analysis of three-dimensional sigma models, augmented by a 't Hooft term for the $U(1)_A$ anomaly,
which represent Landau-Ginzburg-Wilson effective theories for the chiral condensate as the order parameter of
the transition. The~renormalisation group flow based on
the epsilon expansion~\cite{Pisarski:1983ms} predicts the chiral phase
transition to be first-order for $N_f\geq 3$, whereas the case of $N_f=2$ is found to crucially depend
on the fate of the anomalous $U(1)_A$ symmetry: If the latter remains broken at $T_c$, 
the chiral transition should be second
order in the $O(4)$-universality class, whereas its effective restoration would enlarge the chiral symmetry and 
push the transition to first-order. 
For non-zero quark masses, chiral symmetry is explicitly broken.
A first-order chiral phase transition then weakens to disappear
at a $Z(2)$ second-order critical boundary, just like the deconfinement transition in the heavy mass regime, 
whereas a second-order transition disappears immediately. This results in the two scenarios for the Columbia plot depicted
in \fig\ref{fig:columbia}.
The results of the epsilon expansion were essentially confirmed by numerical simulations of three-dimensional sigma 
models~\cite{Gausterer:1988fv} and a perturbative expansion in fixed dimensions~\cite{Butti:2003nu}.
A later high-order perturbative analysis for the case with an effectively restored $U(1)_A$ at the transition temperature 
finds a possibility for second-order transitions also in this case, but~with a symmetry breaking pattern
$U(2)_L\otimes U(2)_R\rightarrow U(2)_V$ signalling a different
universality class~\cite{Pelissetto:2013hqa}. 
Finally, a~recent functional renormalisation group analysis applied to $U(1)_A$ restoration in QCD, which is in good 
agreement with lattice susceptibilities at finite quark masses~\cite{Braun:2020ada}, 
favours the $O(4)$ scenario in the chiral limit~\cite{Braun:2020mhk}.

A large number of lattice simulations has been devoted to disentangle which of these situations is realised in QCD.  
When the Columbia plot is considered on the lattice with different lattice spacings, its parameter space is enlarged by
an additional axis. The~chiral critical line then traces out a chiral critical surface whose shape
is discretisation-dependent, while for all valid discretisations it must of course emerge from the same critical line 
in the continuum limit. The~following subsections collect the current evidence on the location of the chiral boundary line
in the Columbia plot.   

\subsection{$\Nf=2$ and $\Nf=2+1$ \label{sec:scale}}

Early simulations (without sophisticated finite size scaling analyses) on coarse $N_\tau=4$ lattices were 
fully consistent with the expected scenarios shown in 
\fig\ref{fig:columbia}: A first-order region could be clearly seen for $\Nf=3$, whereas the smallest 
available masses were consistent with a continuous transition or crossover for $\Nf=2$, both for unimproved
staggered~\cite{Brown:1990ev} and unimproved Wilson~\cite{Iwasaki:1996zt} fermions.
More recently and using finite size scaling of cumulants, 
a narrow first-order region was also identified for $N_f=2$, again for unimproved 
staggered~\cite{Bonati:2014kpa,Cuteri:2017gci}
and unimproved Wilson~\cite{Philipsen:2016hkv} fermions on $N_\tau=4$. However, the~location of the $Z(2)$-boundary
varies widely between these, indicating large cutoff~effects. 

Recent investigations using the Highly Improved Staggered Quark (HISQ) action 
start at the $N_f=2+1$ physical point and then gradually reduce the light quark mass value, until~they correspond 
to $m_{PS} \approx 45$ MeV, on~lattices $N_\tau=6,8,10$ \cite{Ding:2019prx,Kaczmarek:2020err}. 
No sign of a first-order transition is detected.
Thus either a $Z(2)$-critical point bounding a very narrow first-order region is approached, 
or a second-order transition in the chiral limit.   
The analysis employs a renormalisation group invariant combination of chiral condensates as order parameter representing the magnetisation-like variable,
and the light quark masses $m_l=m_{u,d}$ in units of the strange quark mass as symmetry breaking field,
\beq
M=\Delta_{ls}=2(m_s\langle\bar{\psi}\psi\rangle_l-m_l\langle \bar{\psi}\psi\rangle_s)/f_K^4,\quad H=m_{l}/m_s\;.
\label{eq:delta_ls}
\eeq 
Near a critical point the magnetic observables are dominated by universal scaling functions 
\beq
M(t,h)=h^{1/\delta}f_G(z)+\ldots\;, \quad \chi_M(t,h)=\frac{\partial M}{\partial H}=h_0^{-1}h^{1/\delta-1}f_\chi(z)+\ldots\;,
\eeq
with a scaling variable $z=t/h^{1/\beta\delta}$ expressed in terms of the reduced temperature and external field, 
$t=t_0(T-T^0_c)/T^0_c, h=H/h_0$, which contain the unknown critical temperature in the chiral limit, $T_c^0$, and~
two non-universal parameters $t_0,h_0$.

\begin{figure}[H]
\includegraphics[width=0.36\textwidth]{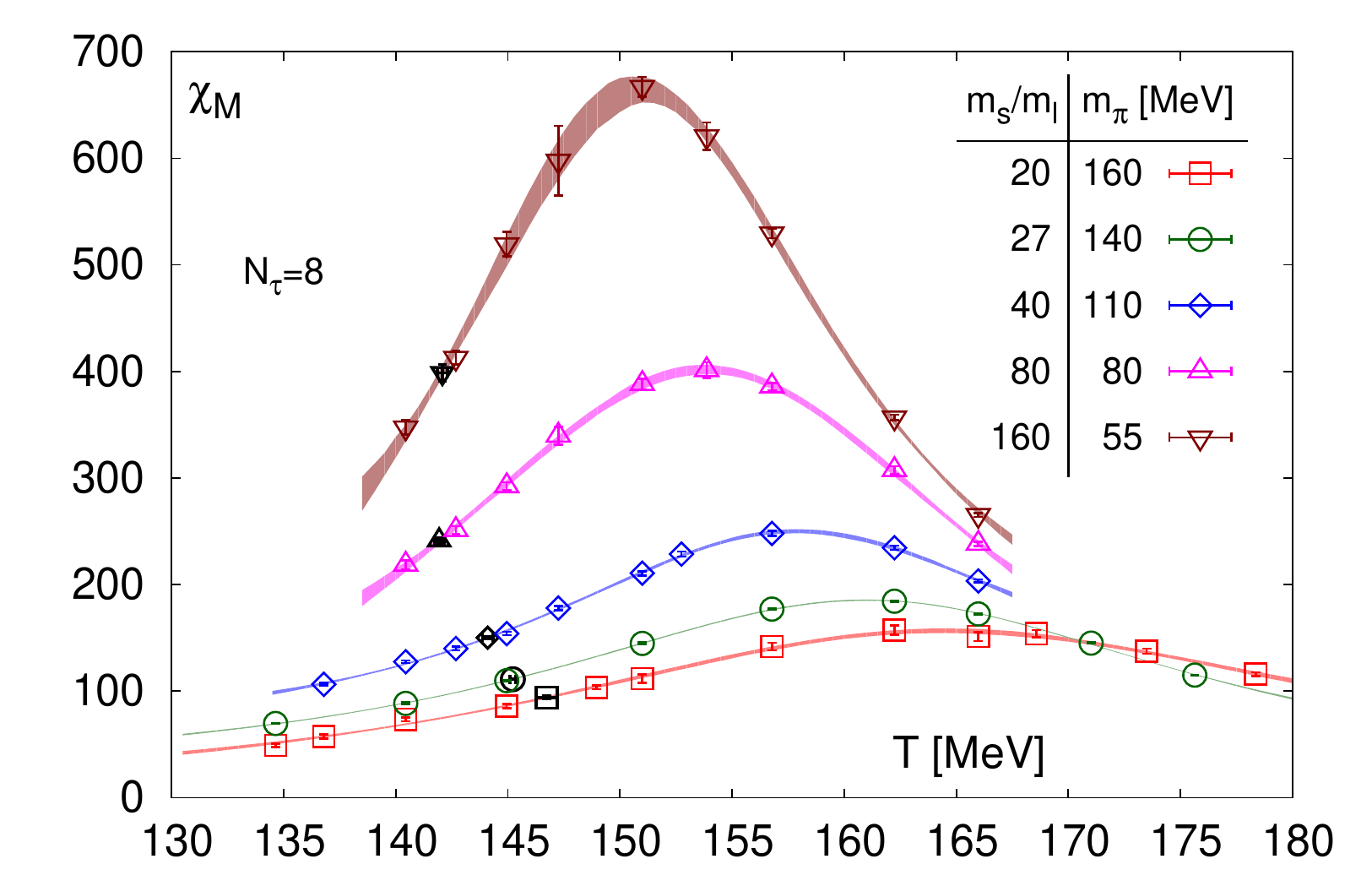}
\includegraphics[width=0.36\textwidth]{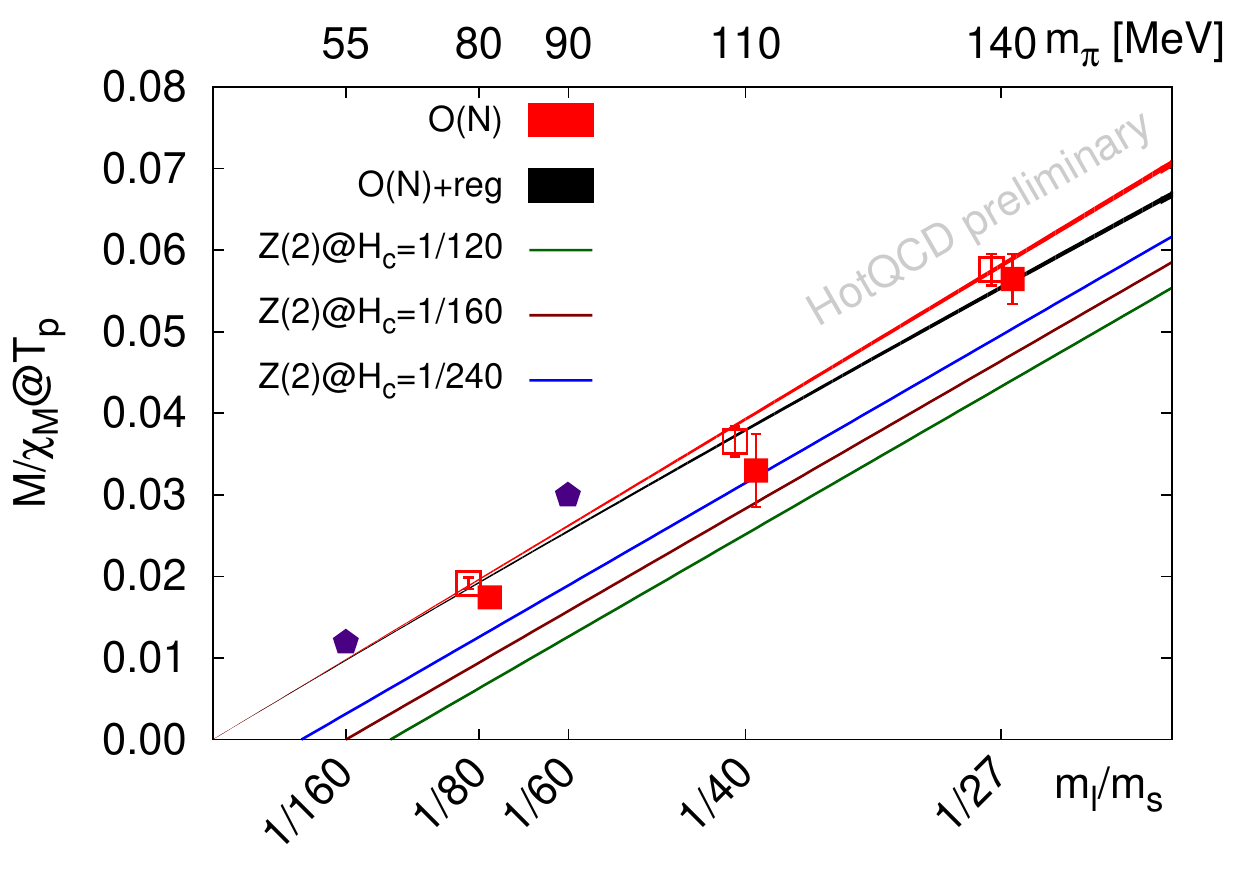}
\caption[]{(\textbf{Left}): Chiral susceptibility at the physical strange quark mass for a range of decreasing light quark masses on
$\NTau=8$ lattices with HISQ action, from~\cite{Ding:2019prx}. 
(\textbf{Right}): Magnetic equation of state at the pseudo-critical temperature 
approaching the chiral limit. Lines represent fits to $O(N)$- ($N=2,4$ second-order scenario) or
$Z(2)$-scaling with a finite critical quark mass (first-order scenario), from~\cite{Kaczmarek:2020err}. 
}
\label{fig:2+1_scaling}
\end{figure}
%

\fig\ref{fig:2+1_scaling} (left) shows the chiral susceptibility as a function of temperature for a series of decreasing
light quark masses on $\NTau=8$. 
A considerable reduction of the pseudo-critical temperature, defined by the peak location of the susceptibility, is apparent as
the light quark mass is reduced. Moreover, the~peak height is growing as expected when approaching a true phase 
transition. To~test for the universality class of the approached transition, 
the ratio $M/\chi_M$ is shown in \fig \ref{fig:2+1_scaling} (right), 
which near a critical point is dominated by universal scaling functions. 
While it is difficult to distinguish between $O(4)$ and $Z(2)$, it is apparent that any finite critical quark mass bounding
a first-order region has to be excessively small to be consistent with the observed behaviour.

There is a similar scaling expression for the approach of the pseudo-critical crossover temperature to the critical
temperature in the chiral limit,
\bea
T_{pc}(H)&=&T_c(m_l=0)\Big(1+\frac{z_X}{z_0}H^{1/\beta\delta}\Big) + \mbox{sub-leading}\;, \\ 
T_c(m_l=0)&=&132^{+3}_{-6}\; \mathrm{MeV}\;.
\label{eq:tc}
\eea
In~\cite{Ding:2019prx} the variation between the possible sets of critical exponents is observed to be very small, so that an
extrapolation makes sense even without knowledge of the true universality class.
Moreover, the~extrapolations were checked to be
stable under an exchange of the order of the continuum and chiral extrapolations, leading to the critical temperature 
as a first result for the chiral phase transition in the massless~limit.

\begin{figure}[H]
\includegraphics[width=0.47\textwidth]{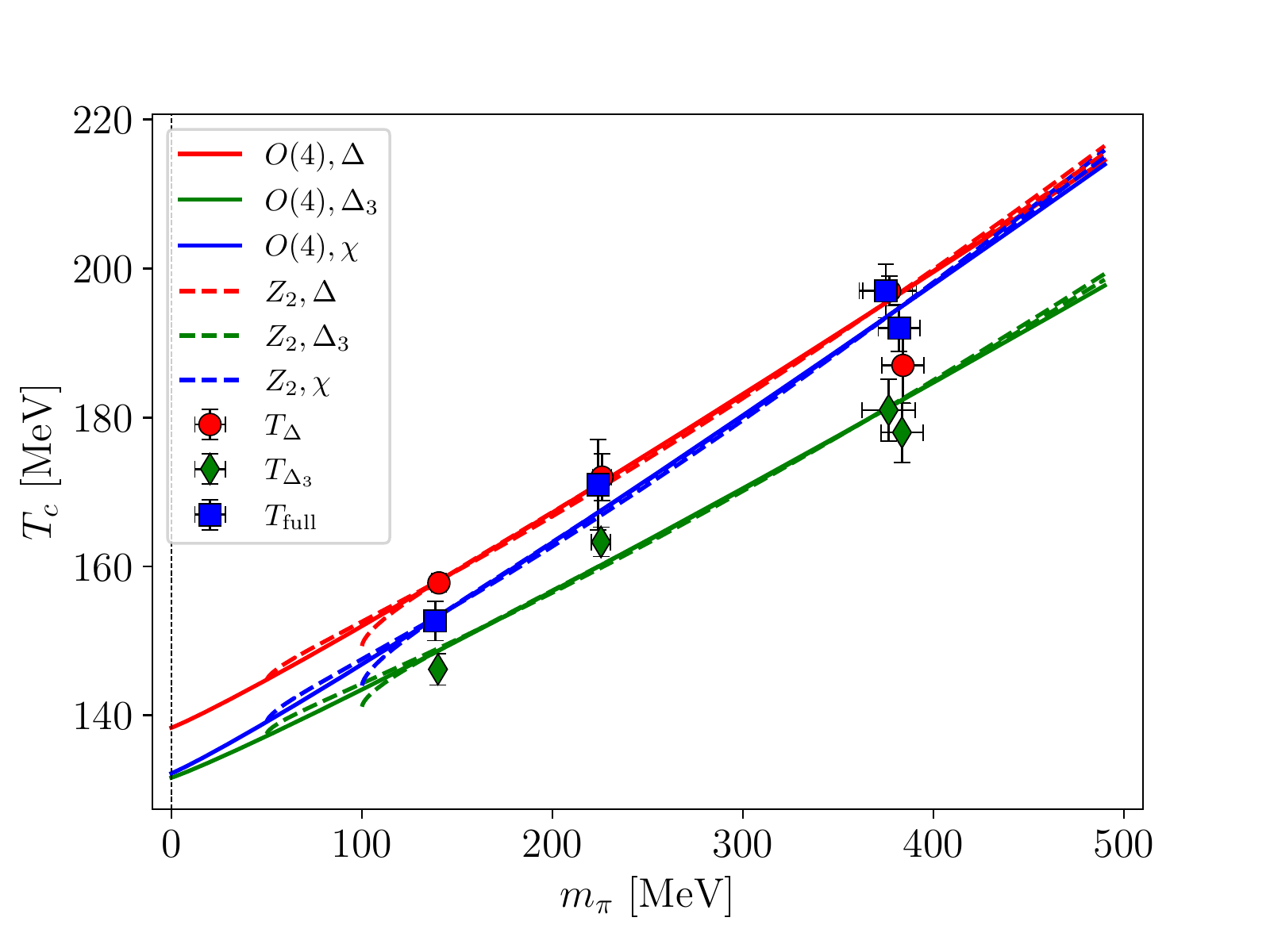}
\includegraphics[height=0.43\textwidth]{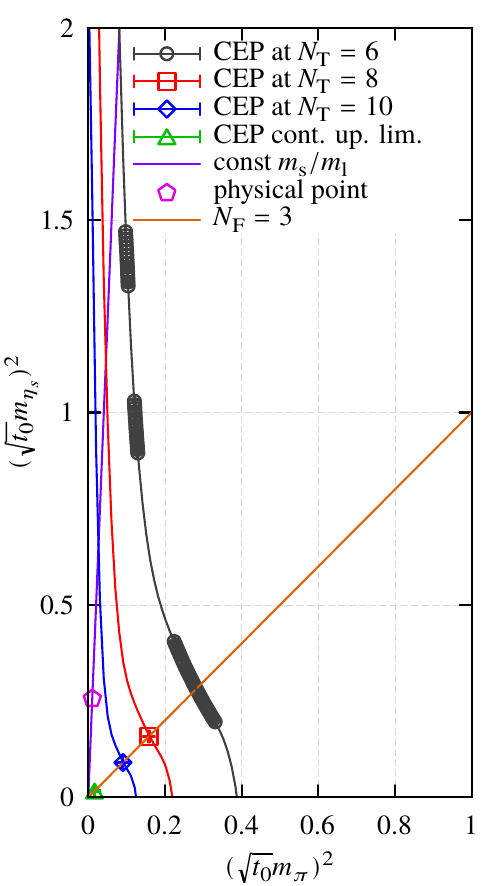}
\caption[]{(\textbf{Left}): Pseudo-critical temperature of the crossover defined by the chiral susceptibility $\chi$, the~inflection
point of the chiral condensate $\Delta$ or an additively renormalised chiral condensate $\Delta_3$, for~$\Nf=2+1+1$ 
twisted mass Wilson fermions close to the continuum. Lines represent chiral extrapolations according to the $O(4)$ second-order
or finite critical $Z(2)$-mass scenario. From~\cite{Kotov:2021rah}. (\textbf{Right}): Columbia plot expressed in $\eta,\pi$-masses in 
units of the Wilson flow parameter $t_0$. Critical points have been determined using an $O(a)$-improved Wilson action.
The first-order region includes the physical point on coarse lattices, but~shrinks drastically as $\NTau$ is increased. 
From~\cite{Nakamura:2019gyy}.  
}
\label{fig:nf211}
\end{figure}

Simulations with Wilson fermions do not yet reach such small quark masses, but~are fully consistent
with this picture down to the physical pion mass. 
In particular, extrapolations of the pseudo-critical temperature to the chiral limit have also been
attempted by an investigation using  $\Nf=2+1+1$ $O(a)$-improved twisted mass Wilson fermions,
with strange and charm quark masses close to their physical values~\cite{Kotov:2021rah}. 
In these simulations lattice spacings are held fixed 
in a range $a/\mathrm{fm}\in[0.062,0.082]$, and~temperature is varied by $\NTau\in[12,20]$.
The pseudo-critical temperature is defined in three different ways: By the maximum of the chiral susceptibility $\chi$ 
as above, and~as the inflection point of polynomial fits to the chiral condensate ($\Delta$) and a subtracted version 
without additive renormalisation $(\Delta_3)$. The~resulting temperatures as a function of pseudo-scalar (pion) mass
are shown in \fig\ref{fig:nf211} (left), together with extrapolations to the chiral limit, employing either 
$O(4)$ exponents or $Z(2)$-exponents and a critical pseudo-scalar mass up to $m_\pi\sim 100$ MeV. Again, it is not
possible to distinguish between these scenarios. As~in the previous case,  the~extrapolated critical temperature 
in the chiral limit is therefore robust under changes of the critical exponents and quoted as
\beq
T_c ^0=134^{+6}_{-4} \;\mathrm{MeV},
\eeq
in remarkable agreement with the staggered~result. 

\fig\ref{fig:nf211} (right) shows an investigation of sections of the chiral critical line using $O(a)$ clover-improved 
Wilson fermions~\cite{Nakamura:2019gyy}. Starting point are the data for $N_f=3$ to be discussed separately below,
and on $\NTau=6$ further points at larger strange quark masses have been added. The~critical line is then fitted
assuming a tricritical strange quark mass as explained in Section~\ref{sec:tric} plus polynomial corrections.
Note that this discretisation features a much wider first-order region, which even contains the physical point 
on the coarser lattices. This must be a lattice artefact, and~the first-order region rapidly
shrinks as $\NTau$ is~increased. 

Several conclusions can be drawn from these results. Firstly, the~width of a potential first-order region as in 
\fig\ref{fig:columbia} (left) is bounded to a small fraction of the physical light quark (or pion) masses.
Second, the~numerical proximity of the critical exponent combinations $1/(\beta\delta)$  
for the 3D $O(2),O(4)$ and $Z(2)$ universality classes appears to allow for a robust extrapolation of the chiral
transition temperature to the massless limit with remarkably small uncertainties. Conversely this statement
implies, however, that it is impossible to firmly identify the universality class in this way, which would require
exponentially accurate data. This problem might be avoided by looking at the scaling of energy-like variables,
which are governed by the critical exponent $\alpha$ that changes sign between the $O(2),O(4)$ and the $Z(2)$
universality classes. It was shown that the Polyakov loop behaves as an energy-like observable, but~unfortunately
a firm distinction between universality classes would require a further substantial reduction of the light quark mass
~\cite{Clarke:2020htu}. 
Finally, note that the value of $T_c(m_l=0)$ is $\sim 25$ MeV lower than the pseudo-critical temperature at the physical point.
This constrains the possible location of a chiral critical point at finite baryon density, as~will be discussed in Section~\ref{sec:cross}.

\subsection{$\Nf=3$ and $\Nf=4$}

The $\Nf=3$ theory, represented by the diagonal of the Columbia plot, \fig\ref{fig:columbia}, is particularly
interesting because a sufficiently large first-order transition and its associated $Z(2)$-critical endpoint are accessible by 
direct simulations, rather than by extrapolation. However, the~situation has become more
complicated since the early results, as~shown by the compilation of critical pseudo-scalar masses obtained by different actions
and lattice spacings in Table~\ref{tab:mpi}. 
The largest value in the table differs by almost an order of magnitude 
from the lowest bound! Unless any of the employed lattice actions is fundamentally flawed, all of them must eventually
converge to the same continuum limit, which is not in sight yet. However, the~general trend is for the critical
mass to shrink when either a lattice with fixed action is made finer (larger $\NTau$), 
or when improved actions are employed. This points to enormous cutoff effects, which quite generally 
appear to increase the first-order region, i.e.,~make the transition stronger.
Another observation is for the first-order region to be considerably larger in the case of Wilson fermions, which is consistent
with stronger cutoff effects due to the complete violation of chiral~symmetry. 

\begin{specialtable}[H]
\setlength{\tabcolsep}{6.5mm} 
\caption{Summary of previous studies (continued from~\cite{Varnhorst:2015lea,deForcrand:2007rq}) 
of the $N_f=3$ chiral critical point at $\mu_B=0$ .}
\label{tab:mpi}
\begin{tabular}{ccccc}
\toprule
\textbf{Action} & \boldmath{$\NTau$} & \boldmath{$m_{PS}^c$} & \textbf{Ref.} & \textbf{Year} \\ 
\midrule
unimproved staggered & 4 & $\sim 290$ MeV & \cite{Karsch:2001nf} & 2001 \\ 
p4 staggered & 4 & $\sim  67$ MeV & \cite{Karsch:2003va} & 2004 \\ 
unimproved staggered & 6 & $\sim 150$ MeV & \cite{deForcrand:2007rq} & 2007 \\ 
HISQ staggered & 6 & $\lesssim 45$ MeV & \cite{Bazavov:2017xul} & 2017 \\ 
stout staggered & 4-6 & $\sim 0$? & \cite{Varnhorst:2015lea} & 2014 \\ 
\midrule
Wilson-$O(a)$-impr. & 6-8 & $\sim 300$ MeV & \cite{Jin:2014hea} & 2014 \\ 
Wilson-$O(a)$-impr. & 4-10 & $\lesssim 170$ MeV & \cite{Jin:2017jjp} & 2017 \\
Wilson-$O(a)$-impr. & 4-12 & $\lesssim 110$ MeV & \cite{Kuramashi:2020meg} & 2020 \\
\bottomrule
\end{tabular}

\end{specialtable}

\begin{figure}[H]
\widefigure
\includegraphics[width=0.38\textwidth]{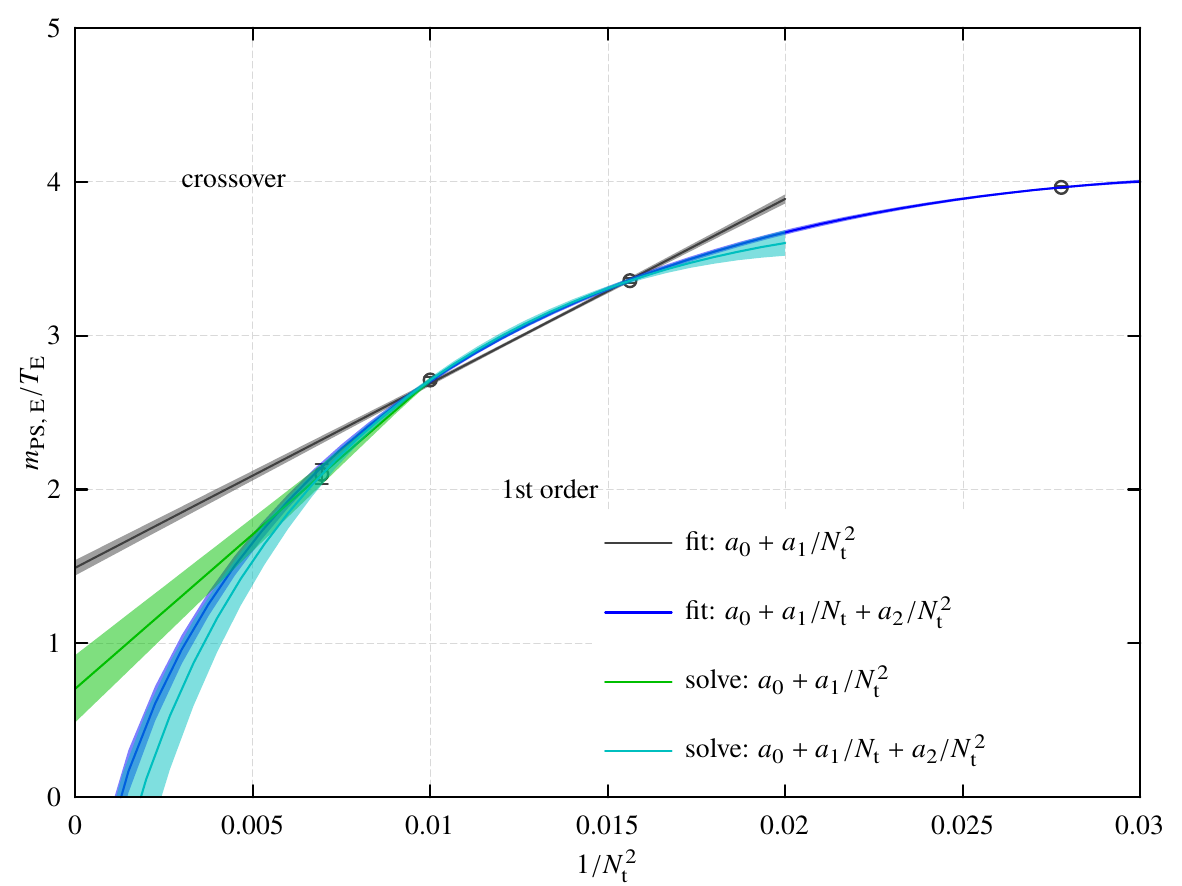}
\includegraphics[width=0.40\textwidth]{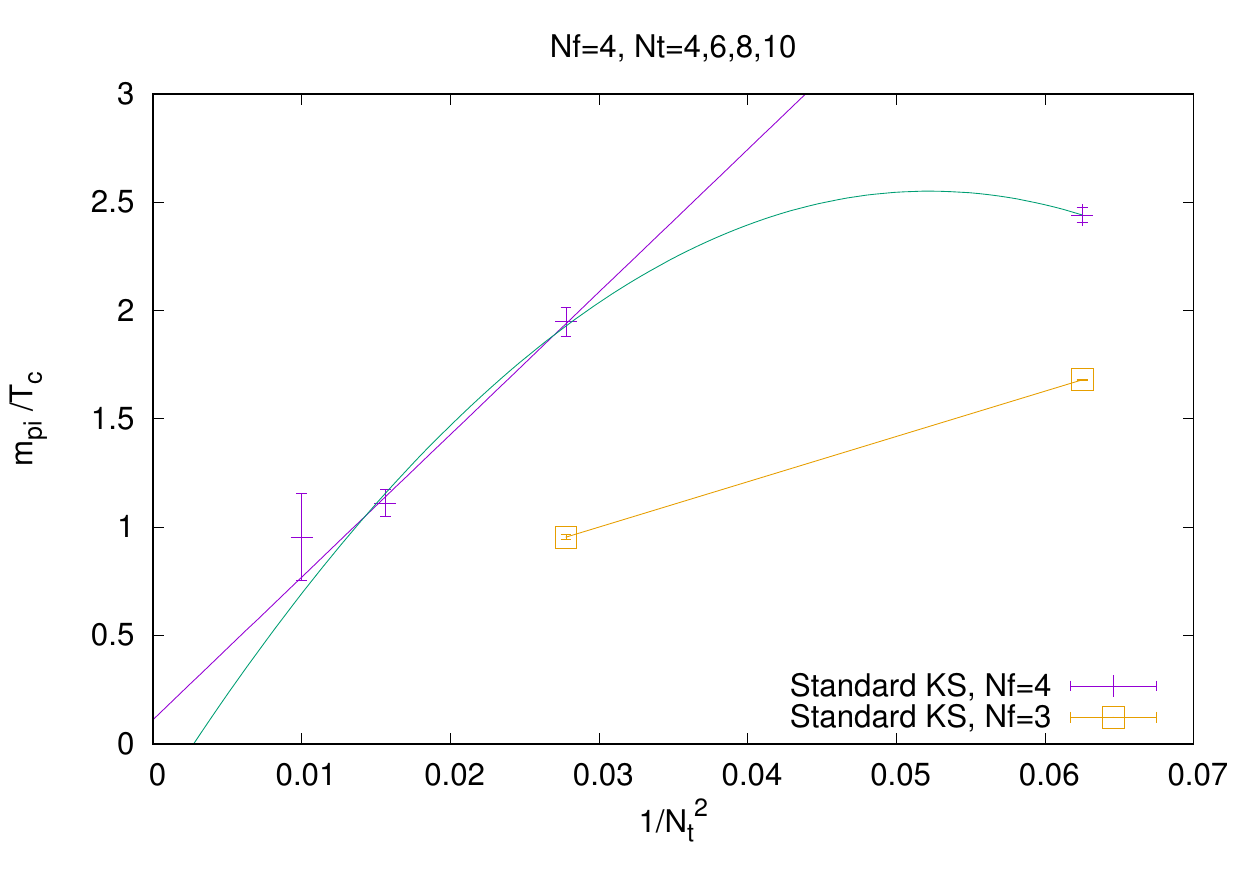}
\caption[]{Critical pseudo-scalar mass in units of temperature as a function of $\NTau^{-2}=(aT)^2$.\\ 
(\textbf{Left}): $N_f=3$ $O(a)$-improved Wilson fermions with
different continuum extrapolations, from~\cite{Kuramashi:2020meg}. (\textbf{Right}): $N_f=3,4$
unimproved staggered fermions, from~\cite{deForcrand:2017cgb}. 
}
\label{fig:nf3-4}
\end{figure}

As an example, consider \fig\ref{fig:nf3-4} ( left), where the critical pseudo-scalar mass is shown from 
a comprehensive long term study with $O(a)$-improved Wilson fermions for different lattice spacings.
The critical meson mass shrinks by nearly a factor of two when the lattice 
is changed between $\NTau=6$ and $\NTau=12$. Its continuum value depends strongly on the extrapolation   
function and could even be zero.
The same effect is observed for $\Nf=4$ unimproved staggered fermions~\cite{deForcrand:2017cgb}, as~shown 
in \fig\ref{fig:nf3-4} (right). One expects the transition in this case to be more strongly first-order than for $\Nf=3$, since the
spontaneously breaking symmetry is larger. This is borne out both for Wilson~\cite{Ohno:2018gcx} and  
staggered fermions. 
Note that for $\Nf=4$ staggered fermions no rooting is applied, which therefore cannot cause any unphysical behaviour. 
Moreover, the~effect of reducing the lattice spacing is similar for $\Nf=2,3,4$ staggered fermions, i.e.,~with and
without~rooting. 

We can then conclude that Wilson and staggered discretisations show the same qualitative behaviour, with~
a monotonic decrease of the critical pseudo-scalar mass $m_{PS}^c(\NTau)$ as the continuum is approached by 
increasing $\NTau$.
Quantitative contradictions between different actions are only avoided, if~the final answer $m_{PS}^c(\infty)$ is below 
the lowest available value $m_{PS}(\NTau)$. In~summary, since 
finer lattices and improved actions have become available, the~situation for $\Nf=3,4$ is quite similarly ambiguous 
to the previously discussed $\Nf=2$ and $\Nf=2+1$.  

%
\clearpage
\end{paracol}
\nointerlineskip
\begin{paracol}{2}
\switchcolumn

\vspace{-12pt}
\subsection{Three-State Coexistence and Tricritical~Scaling \label{sec:tric}}

The uncertainty about the size of a potential first-order region can be removed by a different analysis,
which leads to a consistent picture covering the results of all discreti\-sations reported here.
Starting point is the observation that a change of the chiral phase transition in the massless limit from first order to 
second order, brought about by variation of a continuous parameter, 
necessarily proceeds by a tricritical point~\cite{Rajagopal:1995bc}. 
To see this, assume the scenario in \fig \ref{fig:columbia} (right), 
and consider the corresponding phase diagram for a fixed $m_s<m_s^\mathrm{tric}$, \fig\ref{fig:nf_columbia} (left). 
For non-vanishing quark masses, the~first-order transition weakens to disappear in a critical endpoint, and~the phase diagram is symmetric under a chiral rotation $m_{u,d}\rightarrow -m_{u,d}$.
Hence, in~the massless limit the critical temperature 
marks a triple point of three-state coexistence: Two states with broken chiral symmetry, 
$\pm\langle \bar{\psi}\psi\rangle\neq 0$, and~one with chiral symmetry restored, $\langle \bar{\psi}\psi\rangle=0$. 
When $m_s$ is increased, these points trace out a triple line (corresponding to the green $m_{u,d}=0$ line in \fig\ref{fig:columbia})  
along which the latent heat of the first-order transition
decreases, until~it vanishes in a tricritical point. At~the same time, the~two wings of finite mass first-order transitions
shrink, so that the tricritical point marks the confluence of two critical~endpoints. 
 
This opens an additional path of investigation. Quite generally, tricritical points come with their own set of critical exponents,
which take mean field values since their upper critical dimension is three. 
For a general introduction to the theory of tricritical scaling, see~\cite{lawrie}. 
In particular, the~functional form of the
critical wing lines entering the tricritical point as a function of the symmetry breaking field, which is
the bare light quark mass in our case, is governed by
a critical exponent, 
\beq
\label{eq:scale}
    m_s^c(m_{l})=m_s^\mathrm{tric} + \coeff{a}{1} \cdot m_l^{2/5} + \order[\big]{m_l^{4/5}}\;.
\eeq 
This allows for an extraction  
of the location of a tricritical point with fixed, known exponents, which is much easier than having to determine the
exponents themselves. Knowing the location of the tricritical point then fixes the order of the 
transition on both sides of it, albeit without information about the universality class. 
Attempts in this direction were made in~\cite{deForcrand:2006pv,Nakamura:2019gyy}, but~they are not yet conclusive because 
the lattices are coarse, whereas very high precision data very close
to the chiral limit would be required on sufficiently fine lattices 
in order to unambiguously distinguish tricritical scaling from an ordinary polynomial.
However, this problem can be circumvented by a change of variables in 
the bare parameter space of the theory. 
\begin{figure}[H]
\includegraphics[width=0.35\textwidth]{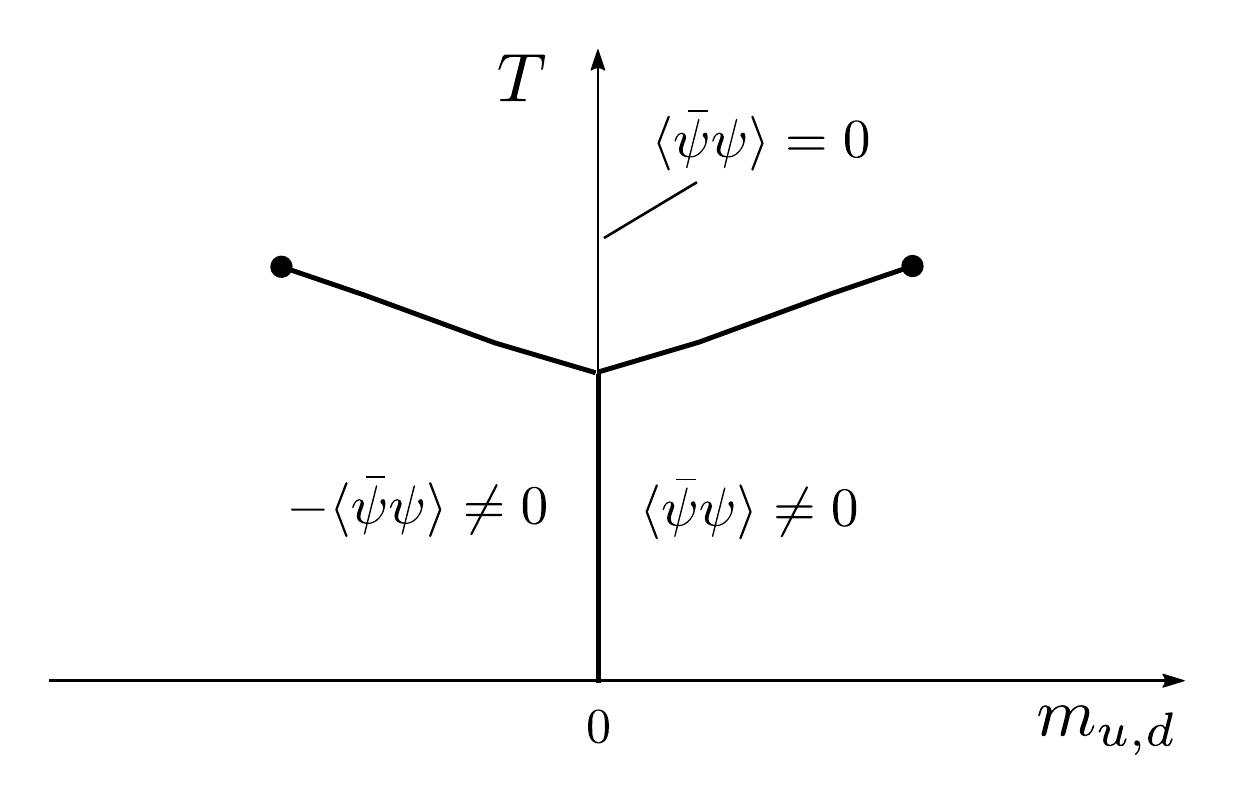}
\includegraphics[width=0.35\textwidth]{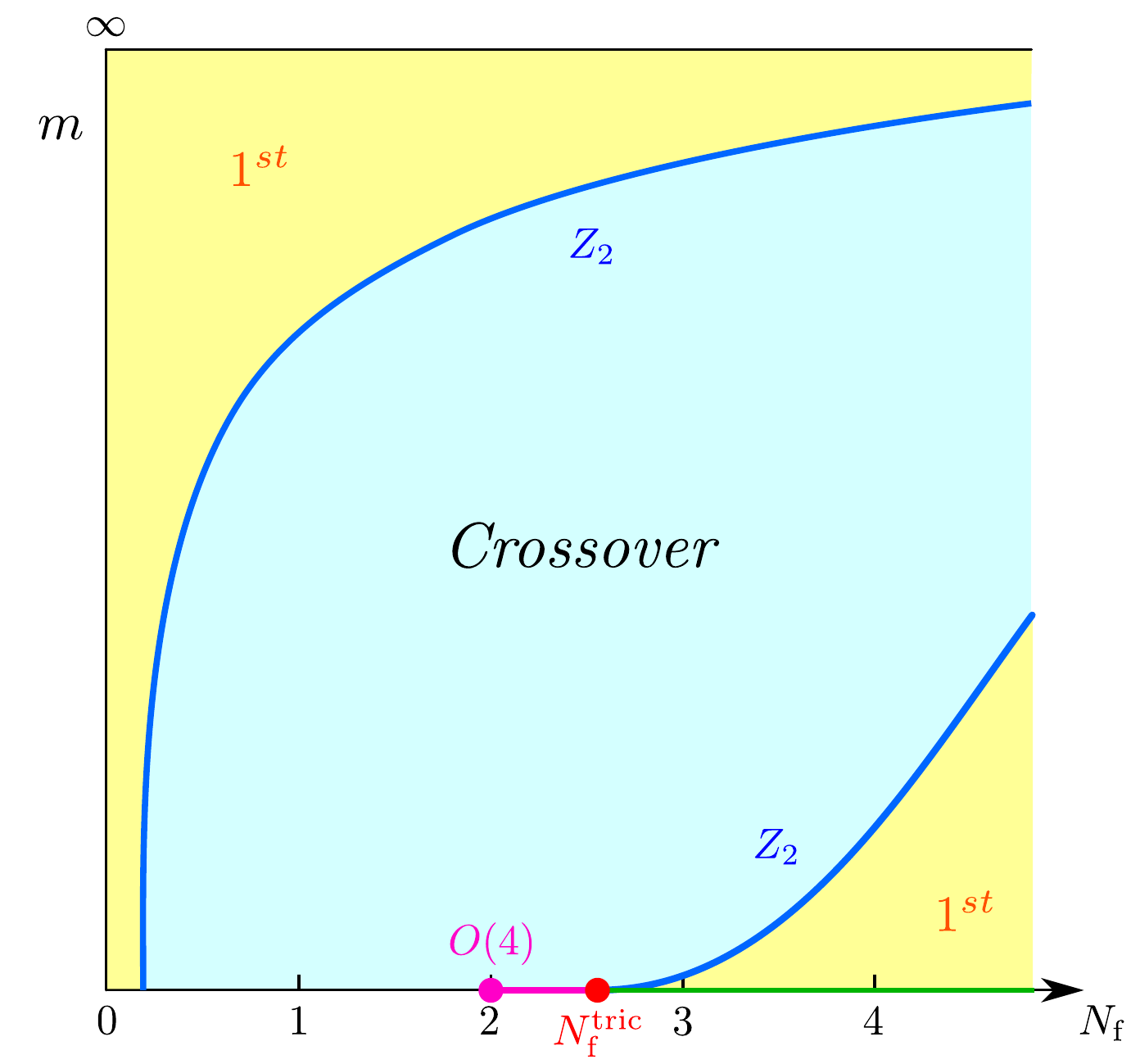}
\caption[]{(\textbf{Left}): Phase diagram with a first-order chiral phase transition. 
For $T<\Tc$ the $T$-axis corresponds to a coexistence line of $\pm\chiralcond\neq 0$, 
and $\Tc(\mud=0)$ represents a triple point.
(\textbf{Right}): Columbia plot for mass-degenerate quarks. Again every point represents a phase boundary 
and has an implicitly associated $T_c(m,\Nf)$. }
\label{fig:nf_columbia}
\end{figure}
\unskip

\subsection{The Chiral Phase Transition as a Function of $\Nf$ and $N_\tau$}

In~\cite{Cuteri:2017gci} it was suggested to consider a different version of the Columbia plot
with degenerate quarks only, in~which the variable strange quark mass $m_s$ is traded for a variable $\Nf$ 
analytically continued to non-integer values. Starting with integer $\Nf$, after~Grassmann integration
the QCD partition function reads   
\beq
Z(\Nf,g,m)=\int {\cal D}A_\mu \; (\det M[A_\mu,m])^{\Nf}\; e^{-\Action_\mathrm{YM}[A_\mu]}\;, 
\eeq
which can now be formally viewed as a statistical system of gauge field variables depending on a continuous parameter $\Nf$.
In the lattice formulation with rooted staggered fermions, whose determinant
is raised to the power $\Nf/4$ in order to describe $\Nf$ mass-degenerate quarks, this is  
implemented straightforwardly. The~
Columbia plot \fig\ref{fig:columbia} (right) then translates to the version shown in \fig\ref{fig:nf_columbia} (right),
where the tricritical strange quark mass is replaced by a tricritical
number of flavours, $2<\Nf^\mathrm{tric}<3$, and~the $\Nf$-axis to the right of it corresponds to the new triple line. 
The crucial advantage in this modified  parameter space is that, 
since there is no chiral transition for $\Nf=1$, a~tricritical point $\Nf^\mathrm{tric}>1$
is guaranteed to exist 
as soon as there is a first-order region for any $\Nf>1$. At~finite lattice spacing there is an additional parameter, 
and the $Z(2)$-critical boundary traces out a chiral critical surface which terminates in a tricritical line in the 
chiral limit of the lattice theory. Since there is ample evidence for first-order transitions at coarse lattice spacings,
the task is now reduced to track their boundaries in the lattice bare parameter space and determine the location 
of the tricritical~line in the lattice chiral limit.  

Numerical results obtained with unimproved staggered fermions for a wide range of flavours and 
$\NTau=4,6,8$ \cite{Cuteri:2017gci,Cuteri:2021ikv} are
shown in \fig\ref{fig:m_nf}.
The left plot represents the lattice version of the Columbia plot \fig\ref{fig:nf_columbia} (right). 
One observes a summary of the previous discussion, namely a first-order transition getting stronger 
with $\Nf$ and with increasing lattice spacing. The~cutoff
effects are growing with $\Nf$, for~$\Nf=7$ the critical bare quark mass is reduced by a stunning factor of five when 
the lattice spacing is reduced ($\NTau$ is increased) by a factor of two! 
No continuum critical mass $m_c$ is discernible yet for any $\Nf$. On~the other hand, the~intercepts of the critical lines
at $m=0$ constitute the sought-after tricritical line $\Nf^\mathrm{tric}(\NTau)$. Tricritical scaling can be appreciated
in the rescaled \fig\ref{fig:m_nf} (right), where the data approach a leading order scaling 
relation~\cite{Cuteri:2017gci}.
This allows for an extrapolation
\begin{align}\label{eq:mscale}
  am_c\big(\Nf(\NTau),\NTau\big)
    &= \coeff{D}{1}(\NTau)\big( \Nf-\Nf^\mathrm{tric}(\NTau)\big)^{5/2}+\ldots \;.
\end{align}
\begin{figure}[H]
\includegraphics[height=4.7cm]{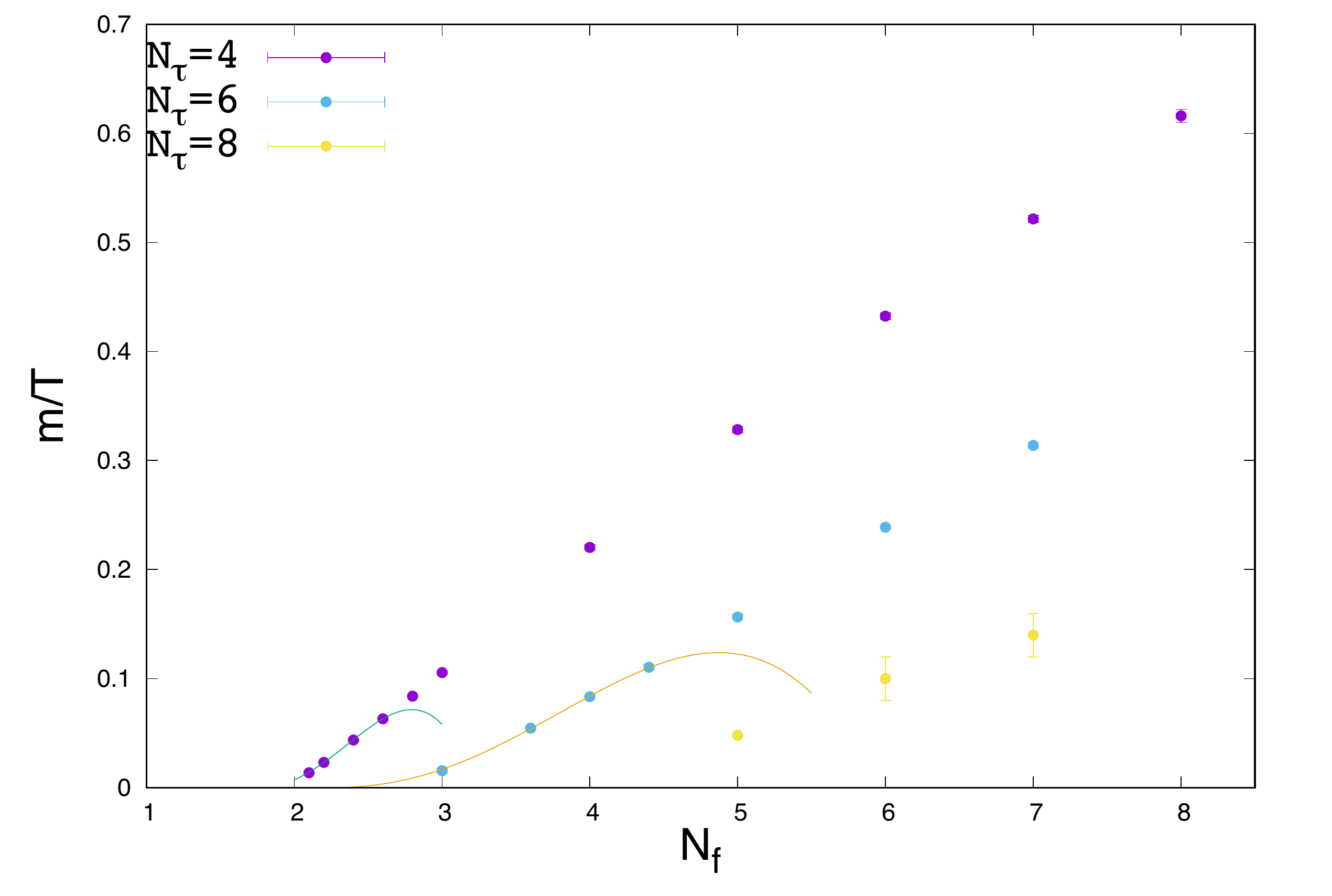}
\includegraphics[height=4.7cm]{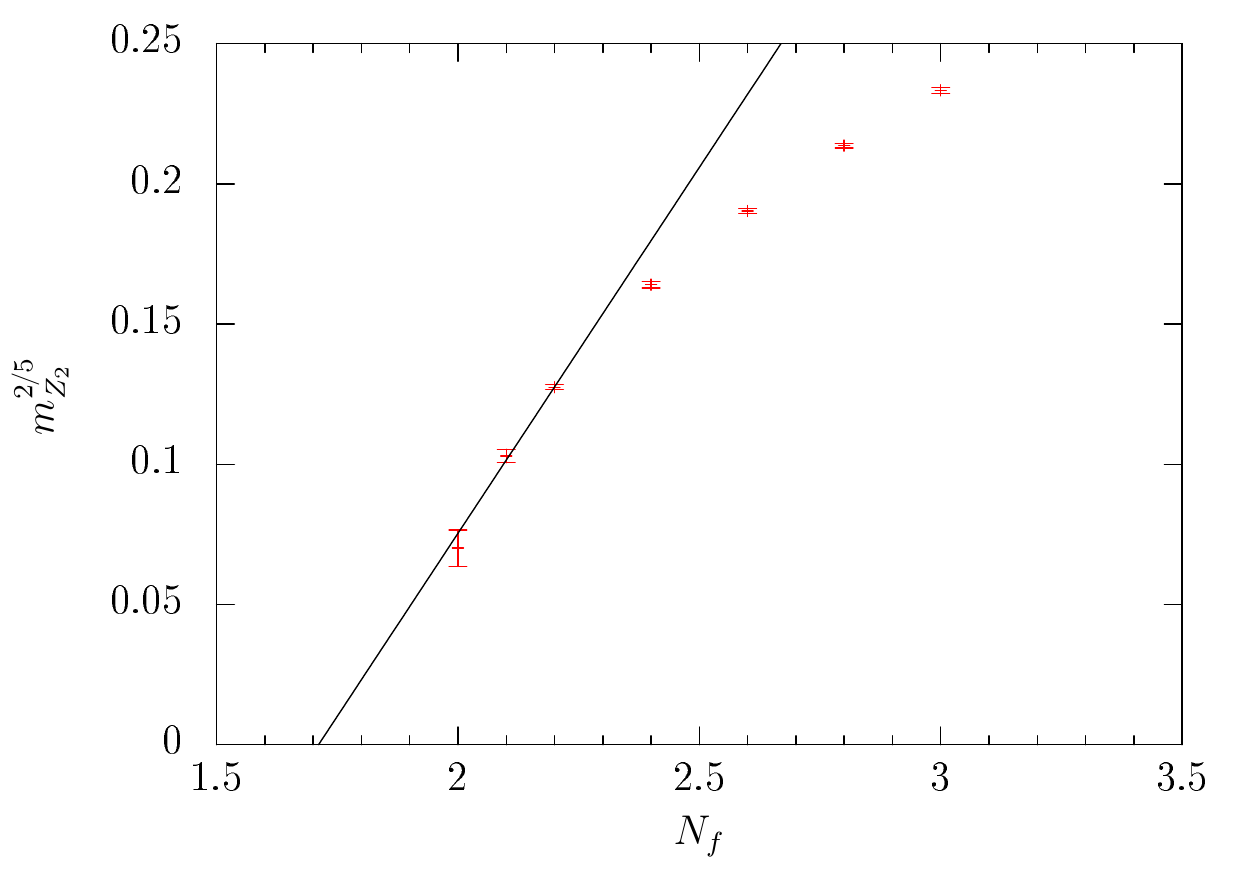}
\caption[]{The chiral critical surface for unimproved staggered fermions, 
projected onto the $(m/T, \Nf)$ plane~\cite{Cuteri:2021ikv} (\textbf{left}) 
and onto the $(am,\Nf)$ plane for $\NTau=4$ \cite{Cuteri:2017gci} (\textbf{right}).
Every point of the plane represents a phase boundary with an implicitly tuned $\beta(am,\Nf,\NTau)$.
Regions above the $Z(2)$-lines represent crossover, those below first-order transitions. 
Fits on the left show next-to-leading order tricritical scaling, on~the right leading-order tricritical scaling.}
\label{fig:m_nf}
\end{figure}
\noindent
 The scaling relation has been inverted here, because~in simulations $\Nf$ is fixed whereas the
 critical quark mass is computed with statistical errors. Note also that the $\Nf=2$ data point
 in  \fig\ref{fig:m_nf} (right) has been obtained by a tricritical extrapolation
 in imaginary chemical potential at fixed $\Nf=2$ \cite{Bonati:2014kpa}. This is an independent confirmation of the 
 bare quark mass as a tricritical scaling field near its chiral limit. Scaling fits in the left figure
 take into account the next-to-leading order terms to predict $\Nf^\mathrm{tric}(\NTau=4)\approx 1.71(3)$ and 
 $\Nf^\mathrm{tric}(\NTau=6)\approx 2.20(8)$. While no extrapolation for $\NTau=8$ is available yet,
 the critical line further flattens and appears to shift to the right, implying a line $\Nf^\mathrm{tric}(\NTau)$.
 One concludes that, for~unimproved staggered fermions, 
 the $\Nf=2$ massless theory shows a first-order transition on $\NTau=4$, but~a second-order transition on all
 finer lattices and in the continuum. The~question now is what happens to the $\Nf\geq 3$ theories.

This can be seen in a different variable pairing in \fig\ref{fig:m_nt} (left), where the same
critical bare quark masses are rescaled by the tricritical exponent, 
and plotted as a function of $\NTau$ for different fixed $\Nf$-values. 
Only a slight curvature is exhibited by those $\Nf$-lines with three data points, which thus are compatible
with next-to-leading order scaling and a tricritical point at some finite $\NTau^\mathrm{tric}(\Nf)$, 
obtained by the corresponding~extrapolations. 

\subsection{Tricritical Scaling for Wilson~Fermions}

What about Wilson fermions, which exhibit the widest first-order region among the discretisation
schemes discussed here? In this case the conceptual situation is more difficult, because~chiral symmetry is broken 
completely so that the bare quark mass and chiral condensate receive additive renormalisation as well as  the 
ordinary multiplicative one. Moreover, the~bare parameter phase diagram is expected to show lattice artefacts
such as parity broken phases~\cite{Aoki:1983qi,Aoki:1986xr} or first-order bulk
transitions~\cite{Sharpe:1998xm,Farchioni:2004us,Burger:2011zc}, whose location depends on the details of the
lattice action and $\NTau$. This may considerably complicate the bare
parameter phase diagram while, of~course, all actions should eventually agree in the continuum~limit. 

Here we consider an RG-improved Wilson gauge and non-perturbatively O(a)-improved Wilson fermion action, 
for which the Aoki phase is limited to the strong coupling regime and no other unphysical phases have 
been observed~\cite{AliKhan:2000wou}, and~which was used for the sequence of $\Nf=3$ simulations~\cite{Jin:2014hea,Jin:2017jjp,Kuramashi:2020meg} displayed in \fig~\ref{fig:nf3-4} (left).
These data have recently been reanalysed in order to test for tricritical scaling~\cite{Cuteri:2021ikv}, 
in analogy to the staggered case above. 
To this end, one can exploit that in chiral perturbation theory 
\beq
m_{PS}^2\propto m_q\;,
\eeq
which also holds for Wilson fermions provided that an additively renormalised quark mass is used. 
The data from \fig\ref{fig:nf3-4} are replotted in  \fig\ref{fig:m_nt} (right), with~the vertical axis rescaled to be proportional
to the renormalised quark mass as scaling variable. Perfect triritical scaling is observed, with~
extrapolations that are robust under variation between leading and next-to-leading order, as~well as when using only 
the coarsest lattices \mbox{$\NTau=4,6,8$}. 
\begin{figure}[H]
\includegraphics[width=0.37\textwidth]{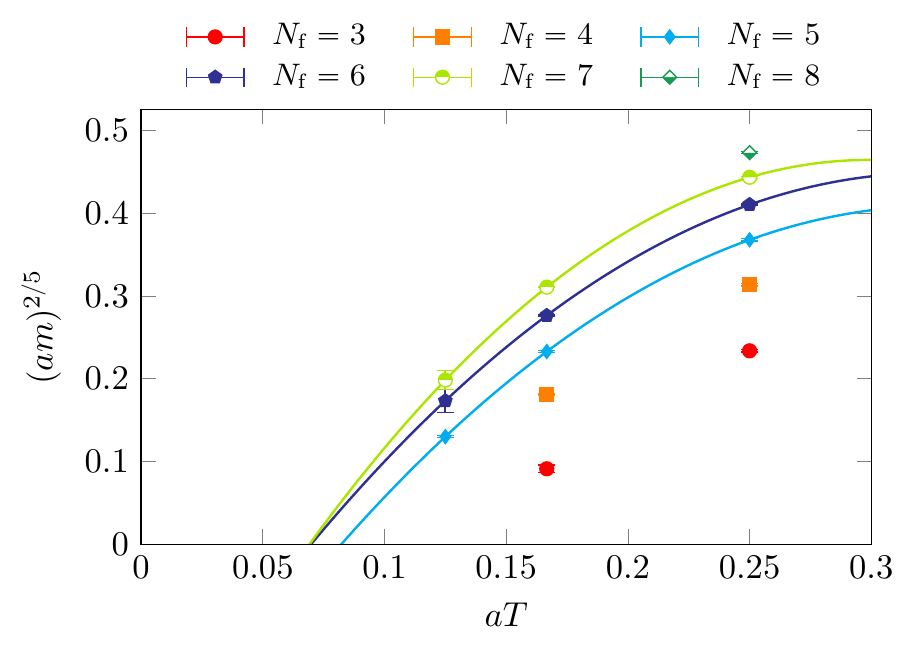}
\includegraphics[width=0.37\textwidth]{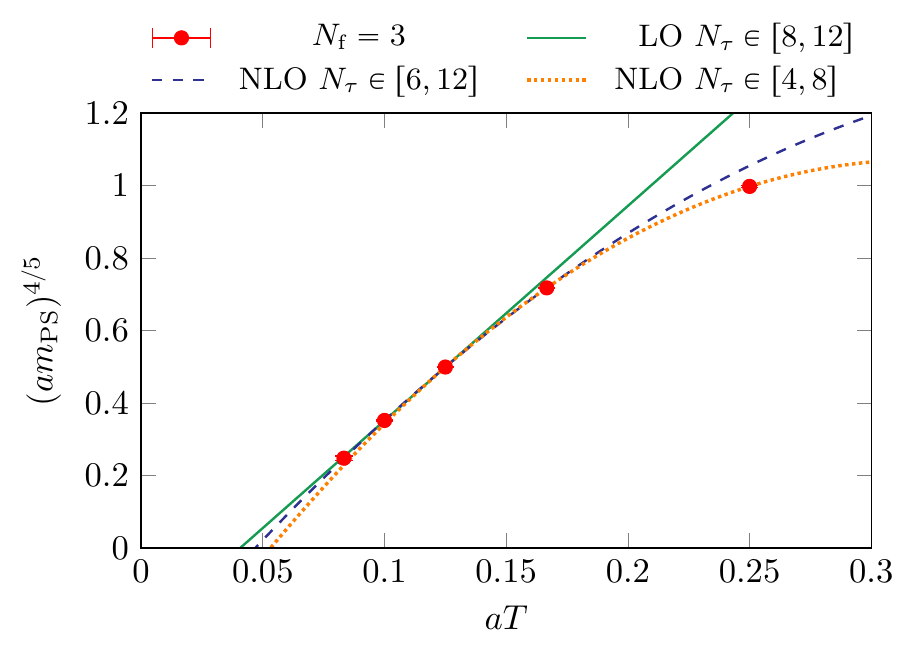}
\caption[]{The chiral critical surface 
projected onto the $(am,aT=\NTau^{-1})$ plane for unimproved staggered fermions~\cite{Cuteri:2021ikv} (\textbf{left})
and for $O(a)$-improved Wilson fermions~\cite{Kuramashi:2020meg} (\textbf{right}), 
with tricritical scaling fits to both.  From~\cite{Cuteri:2021ikv}.}
\label{fig:m_nt}
\end{figure}
\unskip

\subsection{Conclusions for the Continuum~Limit}

Knowing the structure of the bare parameter lattice phase diagram, 
its implications for continuum physics can be stated. The~
continuum limit is represented by the origin of the $(am,\NTau^{-1})$ plane, which can be approached along different lines
of constant physics, representing theories with different values for, e.g.,~the vacuum pseudo-scalar mass. For~any $\Nf$ with 
a finite $\NTau^\mathrm{tric}(\Nf)$, the~parameter region of first-order transitions is not continuously connected to
the continuum limit, and~thus represents a mere cutoff effect. 
In other words, for~all $\NTau>\NTau^\mathrm{tric}$, simulations will only show crossover behaviour for 
\textit{any} vacuum pion mass, such that the transition in the continuum chiral limit is approached from
the crossover, and~corresponds to an isolated second-order point.  
By contrast, a~first-order transition in the chiral continuum limit implies a finite $m_c$ in physical units, which 
would require the $Z(2)$-critical line $am_c(\NTau)$ to terminate in the origin of the plot. 
Moreover, there would be no tricritical point and no associated tricritical scaling, so that  $am_c(\NTau)$ should
have an ordinary Taylor expansion about zero with the usual cutoff corrections $\sim a, a^2$ etc.
In~\cite{Cuteri:2021ikv} it was found that polynomial next-to-leading order fits
are unable to accommodate the $\NTau=4,6,8$ data for $\Nf=5,6,7$ unimproved staggered fermions. For~the $\Nf=3$ $O(a)$-improved Wilson fermions such fits are possible but have 
worse $\chi^2/\mathrm{dof}$  than the tricritical fits.
We must then conclude that the
staggered action predicts all massless
theories with $\Nf\leq 7$ to have second-order transitions in the continuum limit, and~that the Wilson clover-action
agrees with this for $N_f=3$. Note that this conclusion is 
perfectly consistent with the absence of any first-order region in all simulations with improved staggered actions so~far.

In~summary:
\begin{itemize}
\item Qualitatively, the~discretisation
effects on the chiral phase transition are the same for unimproved staggered fermions and either unimproved or 
$O(a)$-improved Wilson fermions,
making the transition stronger compared to the continuum.
\item  In both discretisations, the~boundaries of the first-order transitions at finite lattice spacings exhibit tricritical scaling and extrapolate to a finite $\NTau^\mathrm{tric}(\Nf=3)$. This implies that the first-order region 
is not connected to the continuum limit. 
Thus, when the continuum limit is taken before the chiral limit, as~is necessary to avoid lattice artefacts, 
both predict a second-order transition. 
\item Quantitatively, the~cutoff effects are larger for Wilson fermions, resulting in 
a larger $\NTau^\mathrm{tric}(\Nf)$ than in the case of staggered fermions. This might be explained  by the fact that
the respective discretisations break chiral symmetry fully or only partially.
\end{itemize}

 If these conclusions withstand the test of further investigations, 
 they require a modifi\-cation of the Columbia plot, as~shown in \fig\ref{fig:continuum}, with~ a second-order transition all along the $m_{u,d}=0$ limit, independent of the strange quark mass. Note that 
 the universality class of these second-order transitions remains open. Since the chiral symmetry is different
 for the $\Nf=2,3$ boundary cases, one might expect critical exponents to smoothly cross from one set of values to the other 
 along the second-order~line.
\begin{figure}[H]
\centering
\includegraphics[width=0.4\textwidth]{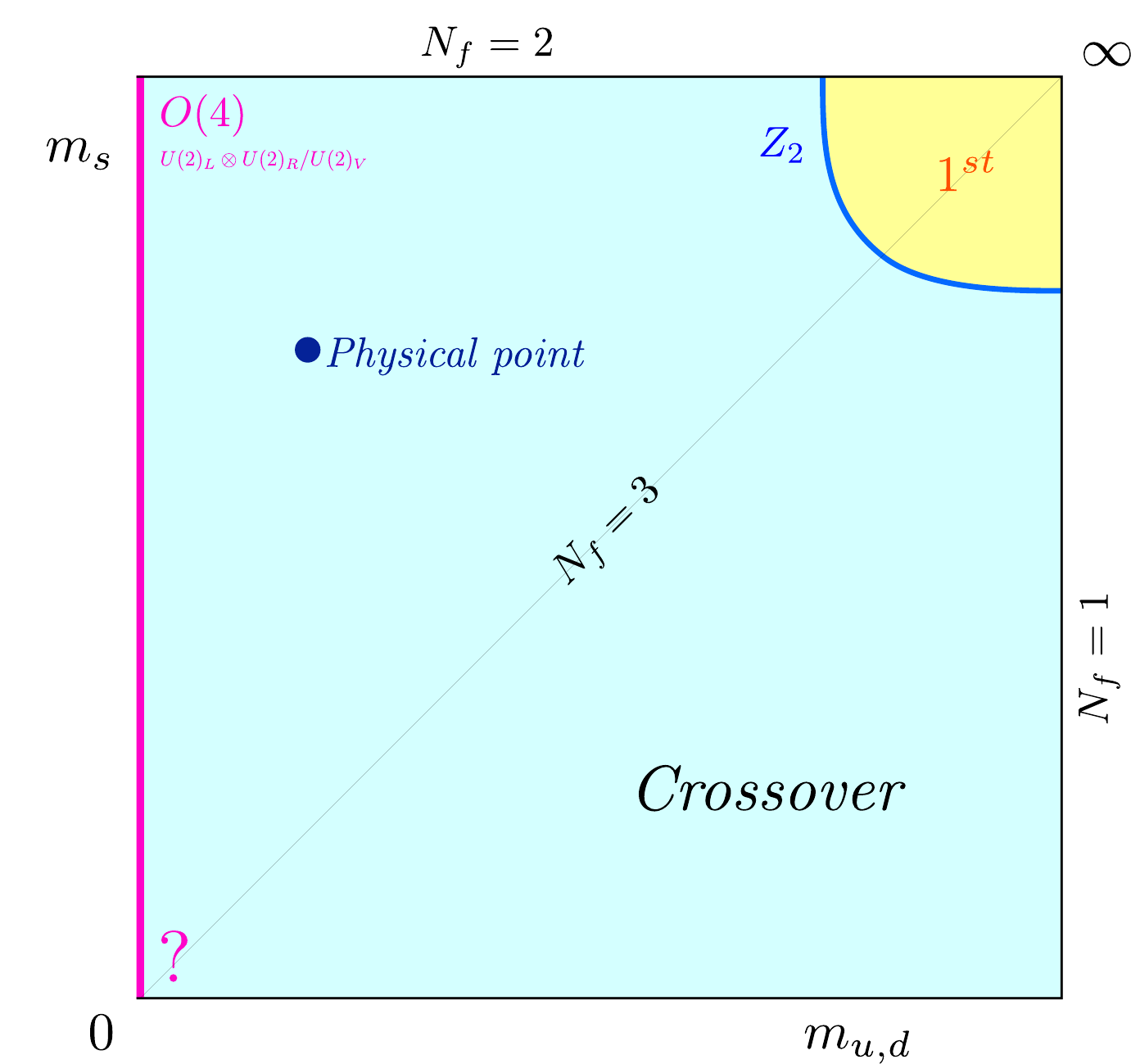}
\caption[]{The Columbia plot in the continuum, as~suggested by the tricritical scaling analyses of unimproved staggered
and $O(a)$-improved Wilson fermions. From~\cite{Cuteri:2021ikv}.}
\label{fig:continuum}
\end{figure}
Numerical simulations are no mathematical proof, and~the reported results are in contradiction with 
the  predictions from~\cite{Pisarski:1983ms,Butti:2003nu}.
Given the presently available data, what would it take to avoid this conclusion and
render the chiral phase transition for $\Nf\geq 3$ first order? This would require 
future data points for $am_c(\NTau)$ (or $am_{PS}(\NTau)$) 
on larger $\NTau$ to extrapolate to zero as a polynomial in $\NTau^{-1}$,
i.e.~\textit{without} tricritical scaling, independent of the discretisation scheme. The~
tricritical scaling observed for the presently available $\NTau$ would then have to be ``accidental''. 
On the other hand, the~question arises
whether the three-dimensional sigma models investigated in~\cite{Pisarski:1983ms,Butti:2003nu} 
represent the most general case compatible with QCD.
It is well-known that $\phi^4$-theories do not permit tricritical points, which requires  
$\phi^6$-terms. In~three dimensions, $\phi^6$ theories are renorrmalisable and moreover show stable infrared fixed
points in non-perturbative studies~\cite{Tetradis:1995br,delDebbio:2011zz}, which could reconcile effective theories
with the current lattice results. It should be possible to resolve these questions in the near future and finally settle the
fate of the thermal QCD phase transition with massless~quarks.

\section{The Columbia Plot with Chemical~Potential}

Once a chemical potential $\mu$ for quark (or $\mu_B=3\mu$ for baryon) 
number is switched on, the~Columbia plot in the continuum
(or at fixed lattice spacing viz.~$\NTau$)
gets extended into a third dimension. In~this case the $Z(2)$-critical boundary lines 
sweep out critical surfaces, as~sketched schematically in \fig\ref{fig:3dcolumbia} (left). Labelling the third axis
by $(\mu/T)^2$, both real and imaginary chemical potential can be discussed.  
Imaginary chemical potential, $\mu=i\mu_i$  with $\mu_i\in \mathbb{R}$, is unphysical, but~it does not induce a sign
problem and thus can be simulated without difficulty~\cite{Alford:1998sd}. This can be exploited to extract various 
properties of the phase diagram 
at sufficiently small real $\mu$ by analytic continuation~\cite{deForcrand:2002hgr,DElia:2002tig}. 
Two exact symmetries are important here. Because~of $CP$-invariance, the~QCD partition function is an even function of the chemical potential $Z(\mu)=Z(-\mu)$. 
Furthermore, for~arbitrary fermion masses
it is periodic in imaginary chemical potential because of the global Roberge-Weiss (or center) 
symmetry~\cite{Roberge:1986mm},
\beq
Z\left(T,i\frac{\mu_i}{T}\right)=Z\left(T,i\frac{\mu_i}{T}+i\frac{2\pi n}{N_c}\right)\;.
\eeq
Thus, as~imaginary chemical potential is increased, the~partition function  
periodically cycles through 
the $N_c$ center sectors, which are distinguishable by the phase of the Polyakov loop, whereas
the thermodynamic functions are invariant under a shift between sectors. The~boundaries of these sectors, located
at $(\mu/T)_c=\pm i(2n+1)\pi /3$ with $n=0,1,2,\ldots$, 
are marked by first-order transitions
for high temperatures and crossover for low tempera\-tures, see \fig\ref{fig:3dcolumbia} (right).
Varying $T$ at fixed imaginary chemical potential, the~analytic continuation of the QCD thermal transition is crossed, 
whose order depends on $N_f$ and the quark masses, as~discussed in the previous sections.
For a first-order chiral or deconfinement transition, the~transition lines
meet up in a triple point, which marks a three-phase coexistence between two phases of the Polyakov loop and their
average.
For a thermal crossover the Roberge-Weiss transition ends in a critical endpoint 
with 3D Ising universality. 
The boundary between these cases, corresponding to specific quark mass values, is marked by a tricritical 
point.
In the Columbia plot for the first Roberge-Weiss plane at $\mu/T=i\pi/3$, these trace out
tricritical lines, in~which the deconfinement critical surface ends in the 
Roberge-Weiss plane~\cite{deForcrand:2010he}.
These structures have been established explicitly 
for unimproved staggered~\cite{deForcrand:2010he,Bonati:2010gi} as well as 
unimproved Wilson~\cite{Philipsen:2014rpa} fermions and generalise to any discretisation respecting the center~symmetry.

\begin{figure}[H]
\vspace*{-0.5cm}
\includegraphics[width=0.37\textwidth]{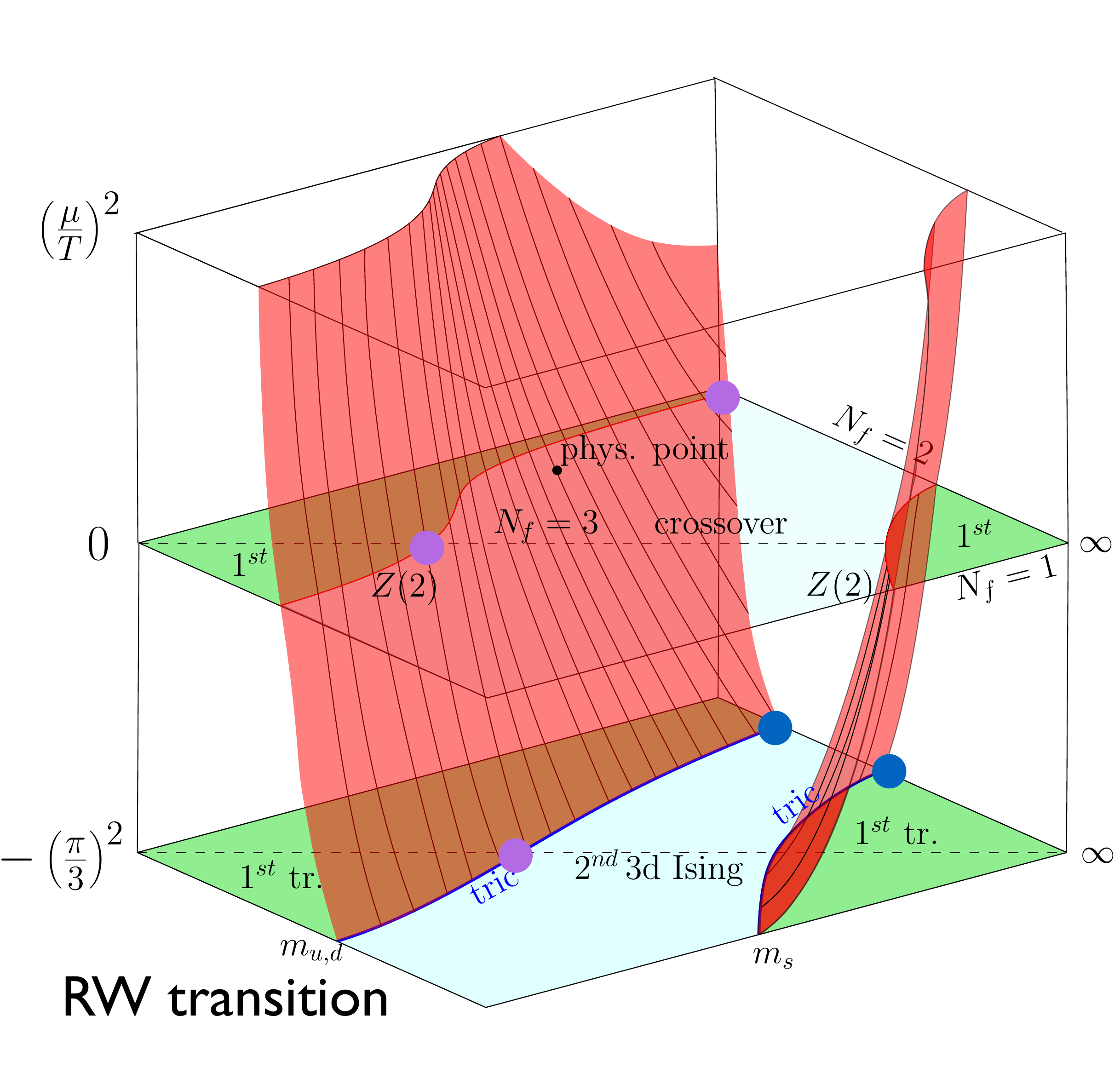}\hspace*{0.5cm}
\includegraphics[width=0.33\textwidth]{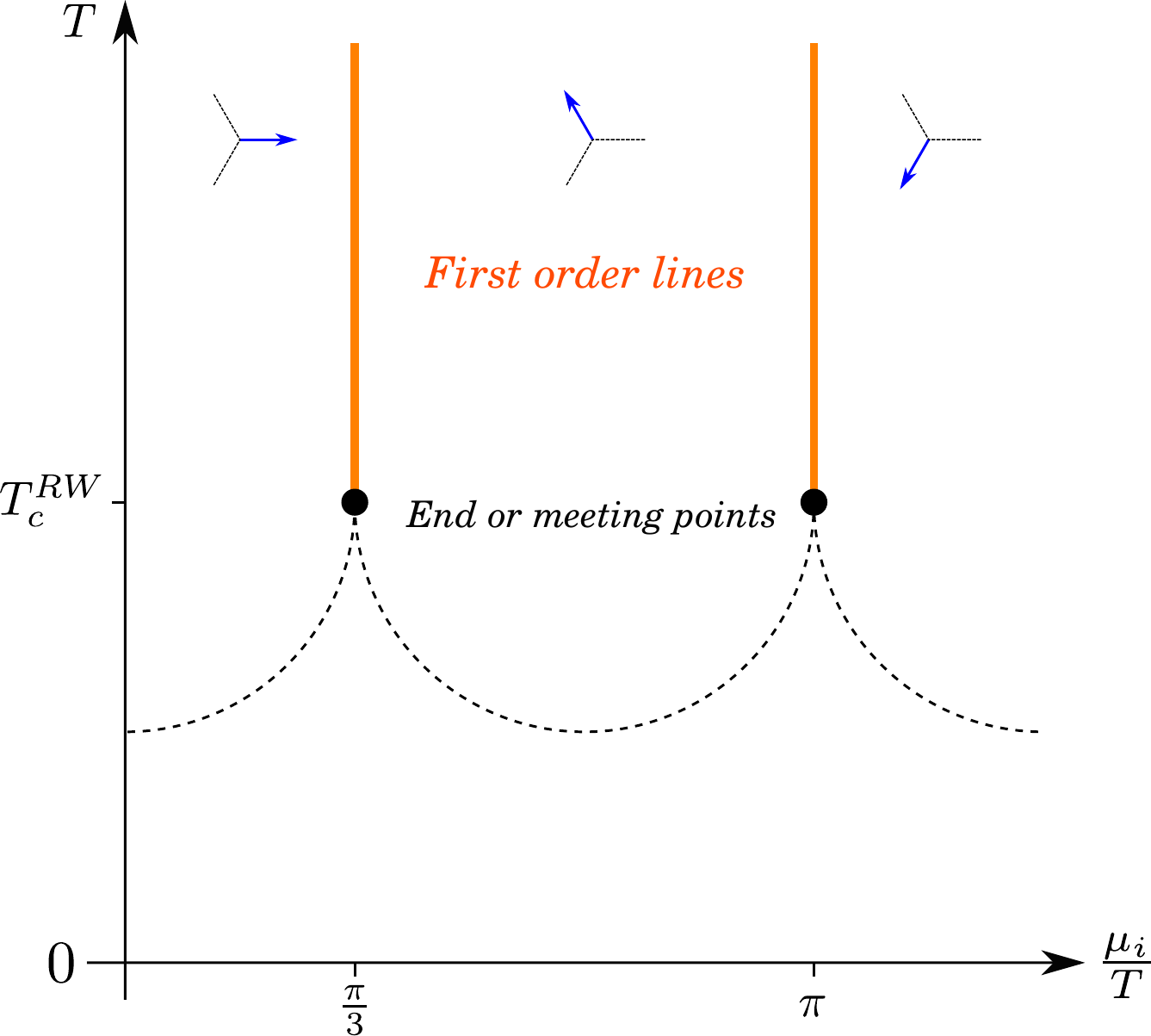}
\caption[]{(\textbf{Left}): Columbia plot with chemical potential as observed on coarse lattices. 
The bottom plane corresponds to the first center-transition.
(\textbf{Right}): The QCD phase diagram at imaginary chemical potential. Vertical lines mark first-order transitions between different
center sectors, the~dotted lines are the analytic continuation of the thermal 
transitions at real $\mu$, whose nature depends on the \mbox{quark masses}.}
\label{fig:3dcolumbia}
\end{figure}

\subsection{The Deconfinement~Transition}

Once again it is instructive to also consider the more easily accessible heavy mass corner, and~to study how the 
deconfinement transition changes in the presence of a baryon chemical potential. Moreover, for~heavy quarks  one
can expand the fermion determinant in inverse quark mass and, after~an additional resummed strong coupling expansion, 
an effective lattice theory can be constructed that reproduces the $Z(2)$-boundary on $\NTau=4$ lattices at $\mu=0$., but~
in addition can be solved at finite chemical potential, both imaginary and real~\cite{Fromm:2011qi}.
The tricritical line in the Roberge-Weiss plane dictates the deconfinement critical surface to emanate from it by tricritical scaling,
\beq
\frac{m_c(\mu)}{T}=\frac{m_c(0)}{T}+K \Big[\Big(\frac{\pi}{3}\Big)^2+\Big(\frac{\mu}{T}\Big)^2\Big]^{2/5}\;,
\eeq
where now the deviation of the chemical potential from the Roberge-Weiss plane is the center symmetry breaking
scaling field. This scaling behaviour is confirmed numerically and, even at leading order, found to extend far into
the range of real chemical potentials. The~region of first-order deconfinement transitions is thus found to continuously 
shrink with real chemical potential~\cite{deForcrand:2010he,Fromm:2011qi}. This observation implies that the chiral and deconfinement transition regions remain
separated for all chemical potentials, i.e.,~the respective critical surfaces do not join. The~chiral critical surface must then 
form a closed volume with the boundaries of the quark mass and chemical potential parameter space, since one cannot
get from a first-order transition to a crossover without passing through a critical~point.

\subsection{The Chiral~Transition}

For unimproved Wilson and staggered discretisations on $\NTau=4,6$, the~3D Columbia plot looks as in  
 \fig\ref{fig:3dcolumbia} (left), with~the region of chiral phase transitions getting wider in the imaginary
 $\mu$ direction. Expanding the $Z(2)$-critical light mass for small chemical potential for any fixed $m_s$,
 \beq
\frac{m^c_l(\mu)}{m^c_l(0)}=1+c_1 \Big(\frac{\mu}{T}\Big)^2+O\Big(\Big(\frac{\mu}{T}\Big)^4\Big)\;,
\label{eq:m_cont}
 \eeq
 one may conclude that $c_1<1$ and by analytic continuation the first order region shrinks in the real-$\mu$ direction.
 As an independent check on these calculations at imaginary chemical potential, the~curvature of the chiral
 critical surface can be computed directly at $\mu=0$. 
 For staggered fermions one finds $c_1<0$ both at $m_s=m_l$ and at $m_s=m_s^\mathrm{phys}$,
 and the next coefficient is negative as well~\cite{deForcrand:2006pv,deForcrand:2007rq,deForcrand:2008vr}.
Thus the chiral transition strengthens with imaginary and weakens 
with real chemical potential. This is opposite to a scenario with a chiral critical point close to the temperature axis, which
would require the chiral transition at the physical point to strengthen with real $\mu$.
The only investigation finding a positive curvature so far is for $O(a)$-improved Wilson fermions~\cite{Jin:2015taa}.
However, we have seen in the previous sections the enormous cutoff effects on the location of the chiral critical surface
at $\mu=0$, and~the question is how these vary with $\mu$ for a given discretisation.
The fact that for $\mu=0$ the critical surface moves to the chiral limit makes it practically impossible to compute 
its curvature on fine~lattices.

Investigations in the Roberge-Weiss plane on finer lattices reveal the same trend as seen 
at $\mu=0$, namely the chiral tricritical line moving towards
smaller quark masses, both for unimproved staggered~\cite{Philipsen:2019ouy} and Wilson~\cite{Cuteri:2015qkq} quarks, 
\fig\ref{fig:rw_scale} (left).
For stout-smeared staggered~\cite{Bonati:2018fvg} and HISQ~\cite{Goswami:2019exb} actions on $N_\tau=4$,
even the larger first-order region in the Roberge-Weiss plane cannot be detected when starting from the physical point and
reducing the pion masses down to $m_\pi\approx 50$ MeV, as~\fig\ref{fig:rw_scale} (right) shows by demonstrating 
second-order scaling for the Roberge-Weiss~endpoint. 
\begin{figure}[H]
\includegraphics[height=3.8cm]{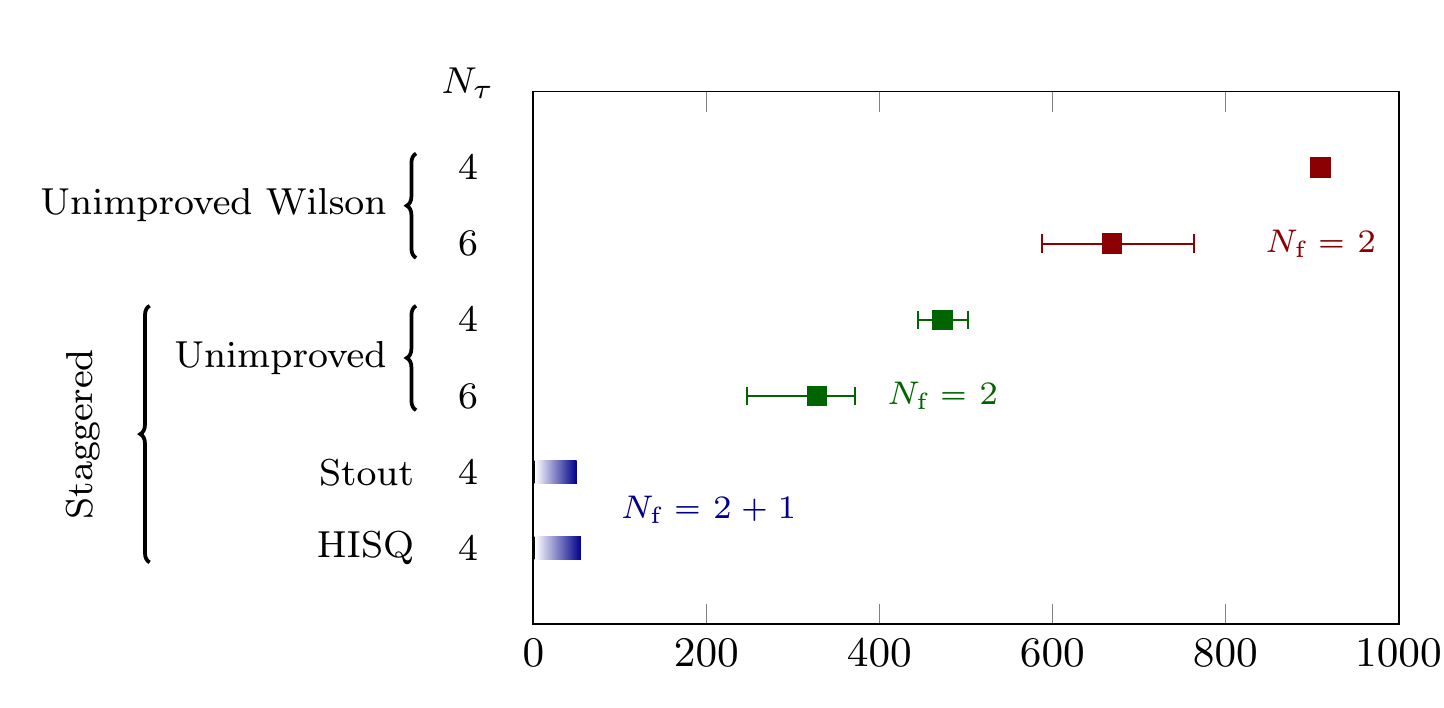}
\put(-35,18){\tiny $m_{PS}^\mathrm{tric}$/MeV}
\includegraphics[height=4cm]{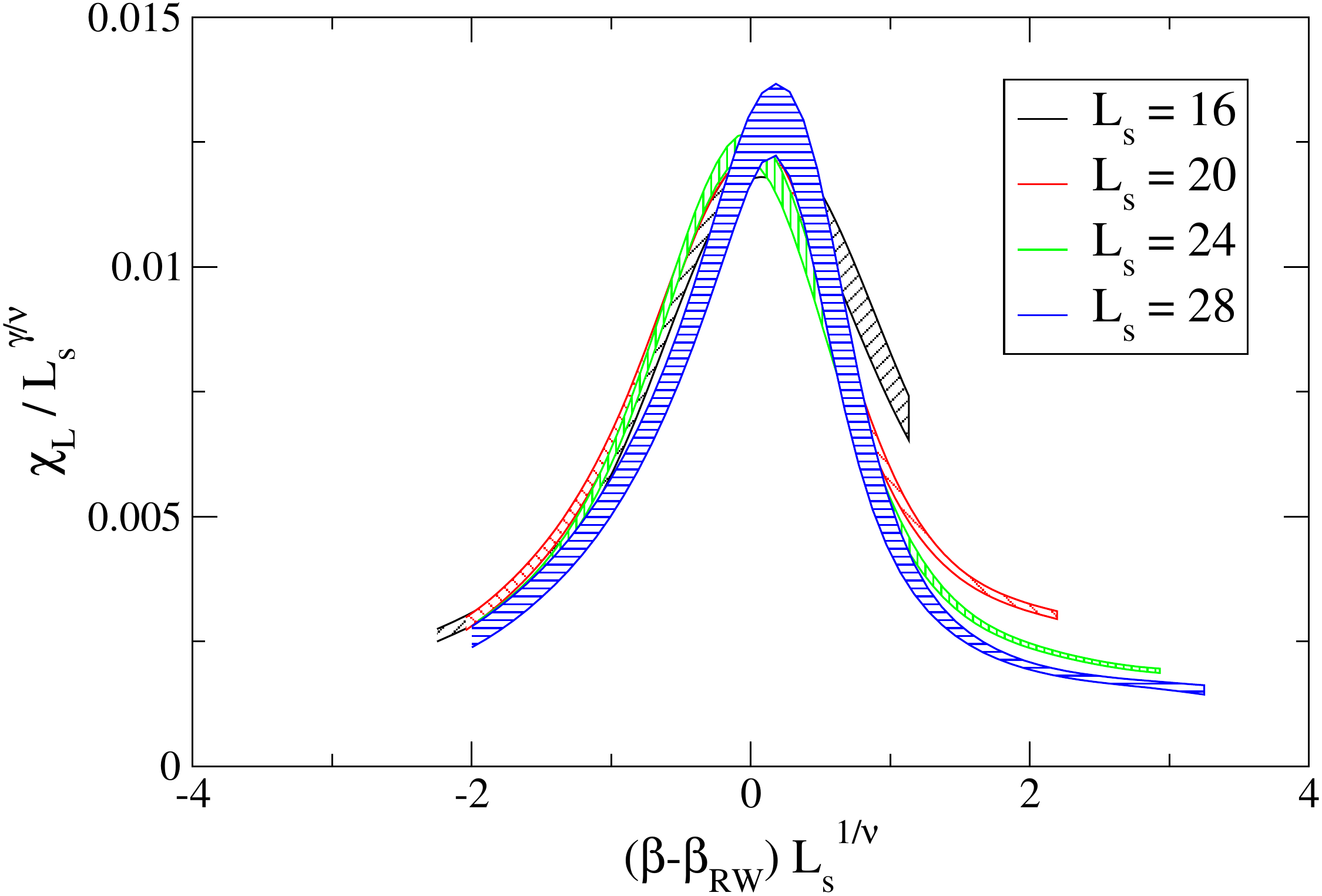}
\caption[]{(\textbf{Left}): Tricritical pseudo-scalar mass values delimiting the first-order 
chiral region in the RW-plane~\cite{Philipsen:2019ouy}. 
(\textbf{Right}): Finite size scaling with 3d $Z(2)$ exponents for stout-smeared staggered fermions with 
$m_{PS}\approx 50$ MeV and $(m_{l}/m_s)_\mathrm{phys}$ on $N_\tau=4$ in the RW-plane~\cite{Bonati:2018fvg}.}
\label{fig:rw_scale}
\end{figure}

Note that the order parameter for the center sectors is the imaginary part of the Polyakov loop, even 
in the light quark regime. It is then interesting to study the interplay between the center and chiral dynamics
in the Roberge Weiss plane, since they are independent symmetries. 
In particular, in~the chiral limit there must be a true phase transition for any chemical
potential, since the latter does not break chiral symmetry. 
\fig\ref{fig:rw_pbp} shows the renormalisation group invariant chiral condensate from Equation~(\ref{eq:delta_ls})
and the disconnected part of the corresponding susceptibility as functions of temperature, as~observed
with HISQ fermions on $\NTau=4$ \cite{Goswami:2019exb}.
Indeed, as~the quark mass decreases
a chiral transition is building up, just as at $\mu=0$, and~it is particularly interesting where the transition happens.
The location of the emerging peak in the susceptibility is, at~the current accuracy, consistent with the yellow bar in 
the figure marking the critical temperature of the Roberge-Weiss endpoint, so that the transition lines in the chiral 
limit would indeed be connected as in \fig\ref{fig:3dcolumbia} (left). Note that there is no a priori reason for this,
since we have independent center and chiral symmetries, with~a $Z(2)$ universality of the Roberge-Weiss endpoint
and a larger, $O(N)$-like universality along the chiral transition line. If~these lines really do meet in the chiral limit,
their junction should be~multicritical. 
%
\begin{figure}[H]
\hspace{-0.5cm}
\includegraphics[width=0.37\textwidth]{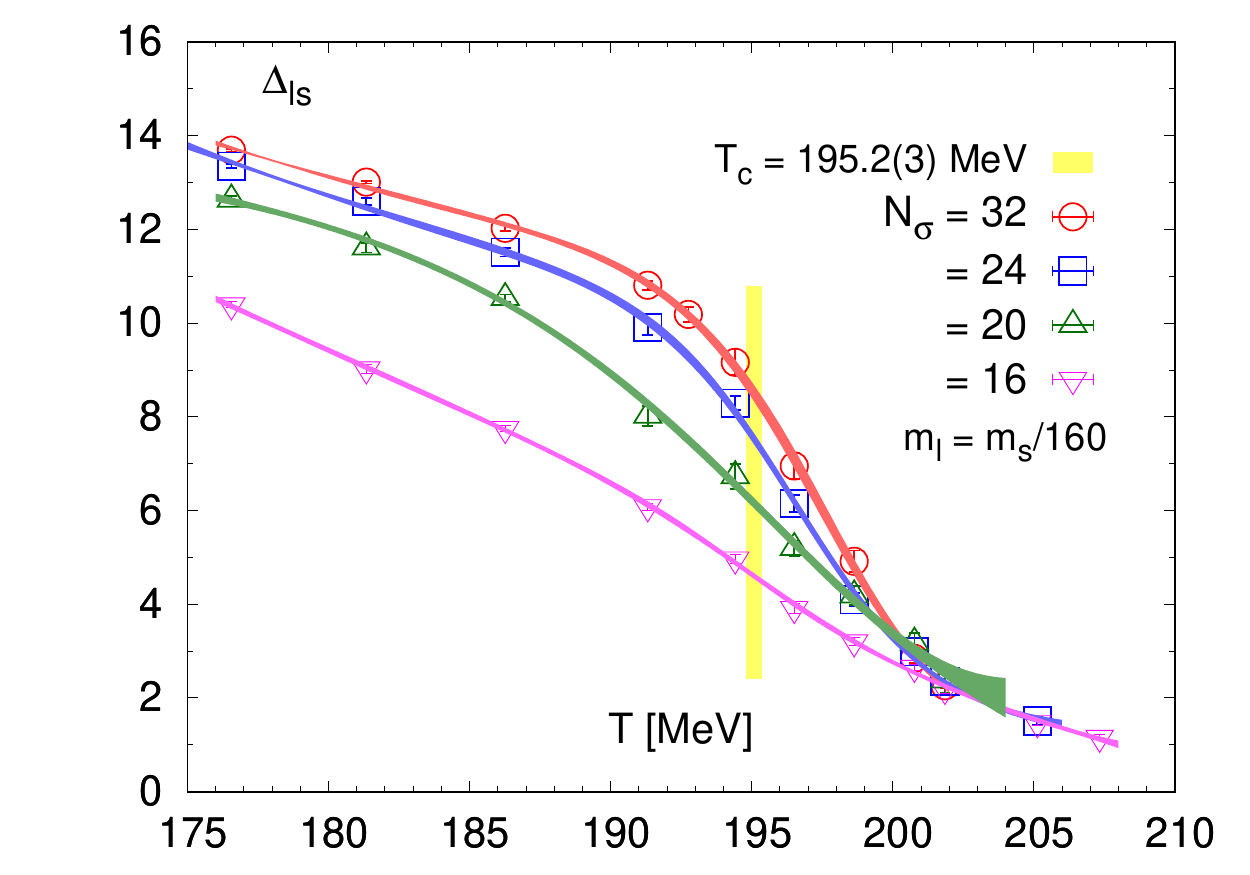}
\includegraphics[width=0.37\textwidth]{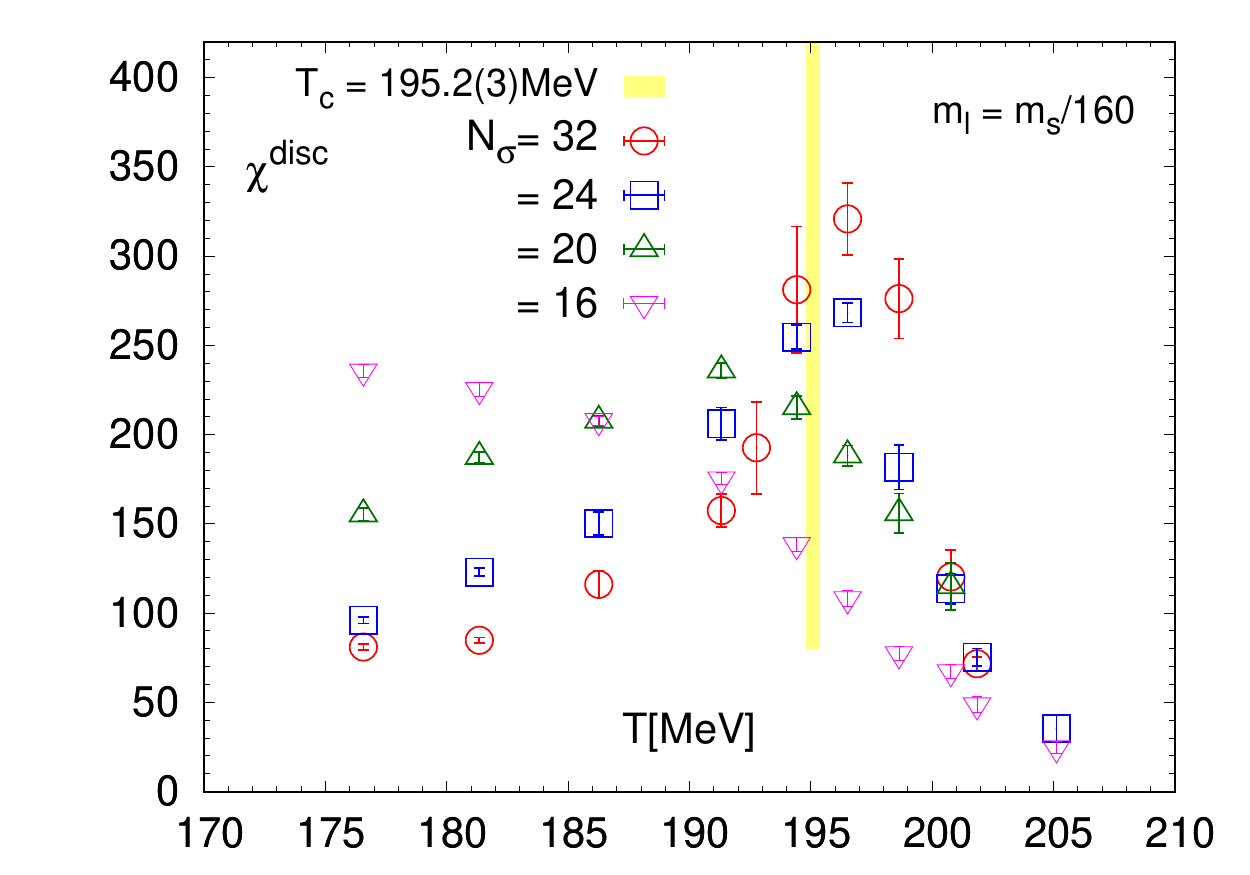}
\caption[]{(\textbf{Left}): Chiral condensate as defined in \eq(\ref{eq:delta_ls}) in the RW plane from HISQ fermions on $\NTau=4$. 
(\textbf{Right}): Disconnected chiral susceptibility at fixed quark mass ratio
for various lattice sizes. Yellow bands show the location of the RW endpoint transition 
temperature, extracted from the peak location of $\chi_L$. From~\cite{Goswami:2019exb}.}
\label{fig:rw_pbp}
\end{figure}
In summary, from~the $\mu=0$ and imaginary chemical potential results together we conclude 
that the entire chiral critical surface is shifting drastically towards 
the chiral limit as the lattice
spacing is decreased, and~it is an open question whether any first-order transition remains in the continuum limit at any $\mu$.
At the same time, this implies that the crossover at the physical point for $\mu/T=\mu_B/(3T)\lsi 1$, which is where an analytic 
continuation of \eq(\ref{eq:m_cont}) holds (see Section~\ref{sec:cross}), is very soft and does not display any singular
behaviour due to the chiral~transition.

\section{QCD at the Physical~Point}

Having discussed the order of the thermal QCD transition in a generalised parameter space, 
the aim is to see how this constrains the physical quark mass configuration.  
The qualitative relation between the phase diagram in the chiral limit of the light quarks and their physical values
has been worked out some time ago by mean field methods~\cite{Halasz:1998qr,Hatta:2002sj},
as shown in \fig\ref{fig:mu_schem}. In~the chiral limit, there must be a non-analytic phase transition for any
value of the baryon chemical potential, since the latter respects chiral symmetry. We have seen that lattice results are
indeed converging towards a second-order transition at $\mu_B=0$. By~contrast, 
the first-order nature of the chiral transition at $T=0$ is based on expectations from NJL, Gross-Neveu and sigma models in 
mostly mean field investigations,
with no firm QCD predictions available yet. If~this assumption holds, there must be a tricritical point marking
the change between first-order and second-order behaviour, which might even have an influence on physical QCD
if the light quark masses happen to be in the scaling window of the tricritical point~\cite{Stephanov:1998dy}. 
At finite quark masses chiral symmetry is broken explicitly, the~second-order line turns into crossover 
and the first-order transition weakens to disappear
in a $Z(2)$-critical line, which emanates from the tricritical point by tricritical scaling~\cite{Hatta:2002sj}, 
\bea
\mu_B^c(m_{u,d})&=&\mu _B^\mathrm{tric}+A \; m_{u,d}^{2/5}+O(m_{u,d}^{4/5}),\nn \\
T_c(m_{u,d})&=&T^\mathrm{tric}+B \; m_{u,d}^{2/5}+O(m_{u,d}^{4/5})\;.
\eea
This implies an ordering of the critical temperatures to be exploited below,
\beq
T_c(m_{u,d}=0,\mu_B=0)>T_ \mathrm{tric}(m_{u,d}=0,\mu_B=0)>T_\mathrm{cep}(m_{u,d}^\mathrm{phys},\mu_B^\mathrm{cep})\;.
\eeq

For completeness, we need to also discuss an alternative scenario, where the chiral phase transition in the massless limit
is second order all the way to $T=0$. At~least from a lattice perspective, this is not excluded so far, but~crucially
depends on whether there is any non-trivial $m^c_{u,d}(\mu)$-dependence in the continuum limit. Moreover, a~recent investigation
of the chiral nucleon-meson and chiral quark-meson models finds the phase transition for $m=0$ at $T=0$ to turn second
order, once fluctuations are included~\cite{Brandes:2021pti}. 
In such a scenario
there is no tricritical point and no first-order transition anywhere.
Instead, non-vanishing quark masses remove the entire second-order line and the chiral transition would be analytic crossover
exclusively for physical quark masses.
\begin{figure}[H]
\centering
\includegraphics[width=0.6\textwidth]{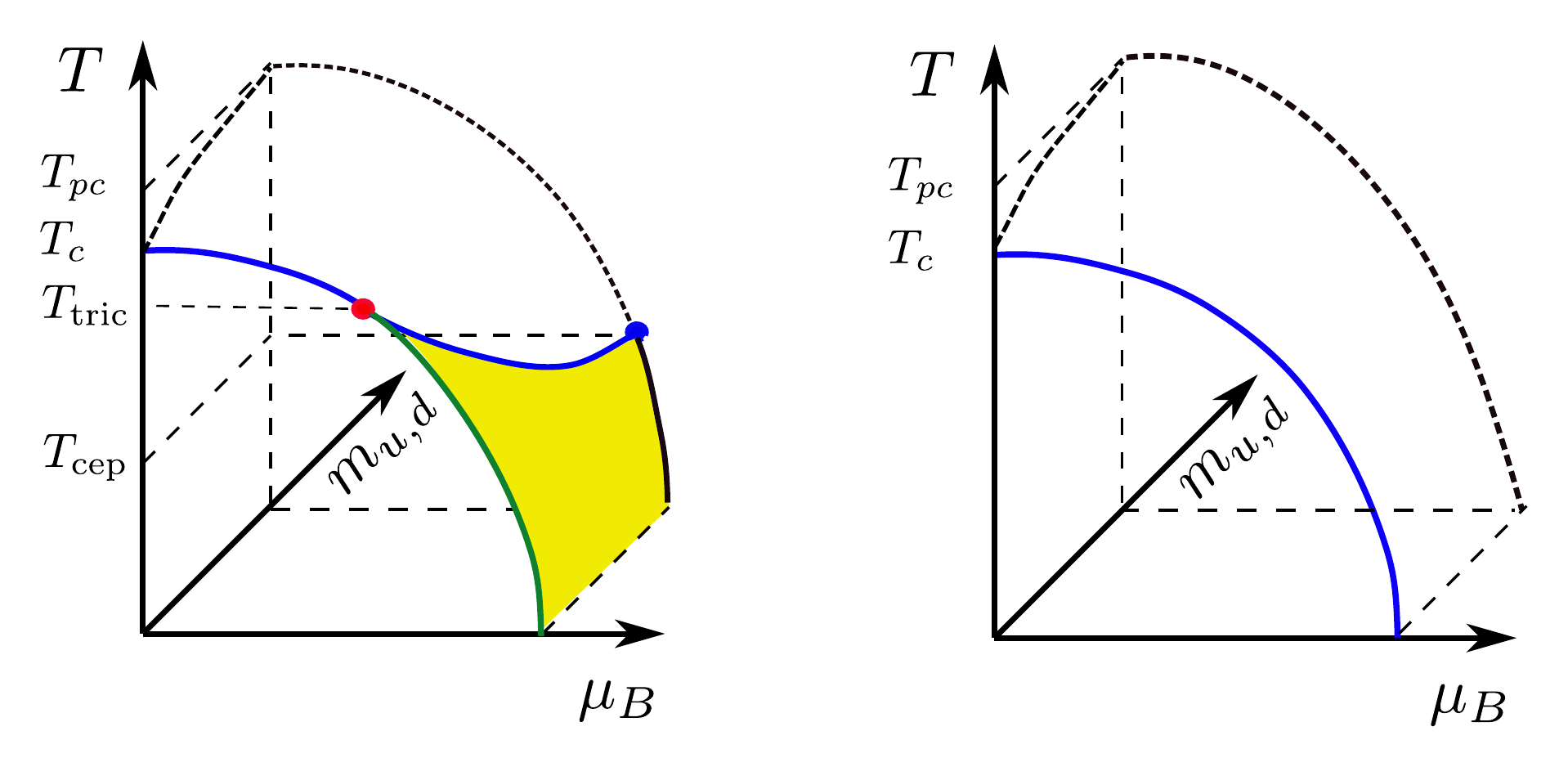}
\caption[]{(\textbf{Left}): Relation of the tentative QCD phase diagram with physical light quark masses 
(back plane) to the chiral limit (front plane) according to~\cite{Halasz:1998qr,Hatta:2002sj}. (\textbf{Right}): If
the entire chiral transition line in the massless limit is of second order, the~transition at the physical point is
crossover everywhere. }
\label{fig:mu_schem}
\end{figure}
%

\subsection{The Crossover at Small Baryon~Densities \label{sec:cross}}

There are several methods that have been used so far to extract information about
the phase structure at the physical point for small baryon density. \textls[-15]{All of them introduce 
some approximation which can be controlled 
as long as $\mu/T\lsi 1$: (i) Reweighting~\cite{Fodor:2001au}, \mbox{(ii) Taylor} expansion in $\mu/T$ \cite{Allton:2002zi} and (iii) analytic continuation from imaginary chemical \mbox{potential~\cite{deForcrand:2002hgr,DElia:2002tig}}. 
When the QCD pressure is expressed as a series in baryon \mbox{chemical potential}},                
\beq
\frac{p(T,\mu_B)}{T^4}=\frac{p(T,0)}{T^4}+\sum_{n=1}^\infty \frac{1}{2n!} \chi^B_{2n}(T) \left(\frac{\mu_B}{T}\right)^{2n}\;,\quad 
 \chi^B_{2n}(T)=\frac{\partial^{2n} (\frac{p}{T^4})}{\partial (\frac{\mu_B}{T})^{2n}}\Big|_{\mu_B=0}\;,
 \label{eq:press}
\eeq
the Taylor coefficients are the baryon number fluctuations evaluated at zero density, which can also be computed by fitting
to untruncated results at imaginary $\mu_B$. This permits full control of the systematics between (ii) and (iii).
These coefficients are presently known up to $2n=8$ on $N_\tau=16$ lattices, \fig\ref{fig:cemtest} (left), and~
in principle also observable experimentally.
For a review of the equation of state relating to heavy ion phenomenology, see~\cite{Karsch:2015nqx,Ratti:2019tvj}. 
Note also, that this low density regime
appears to be accessible by complex Langevin simulations without recourse to series expansions, 
albeit not yet for physical quark masses~\cite{Attanasio:2020spv}. This offers an additional cross check between different
methods.

\begin{figure}[H]
\vspace*{-0.5cm}
\includegraphics[width=0.34\textwidth]{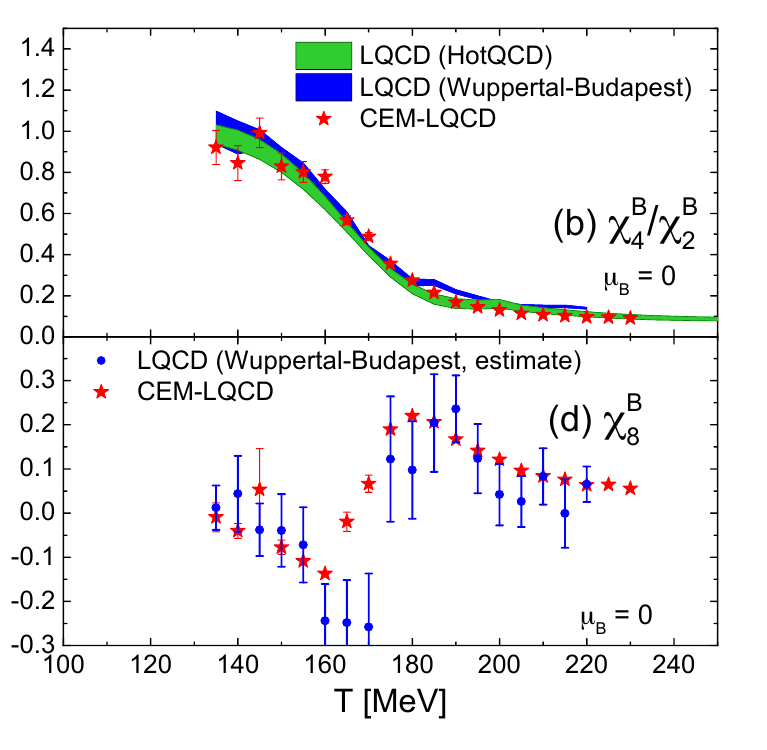}
\includegraphics[width=0.37\textwidth]{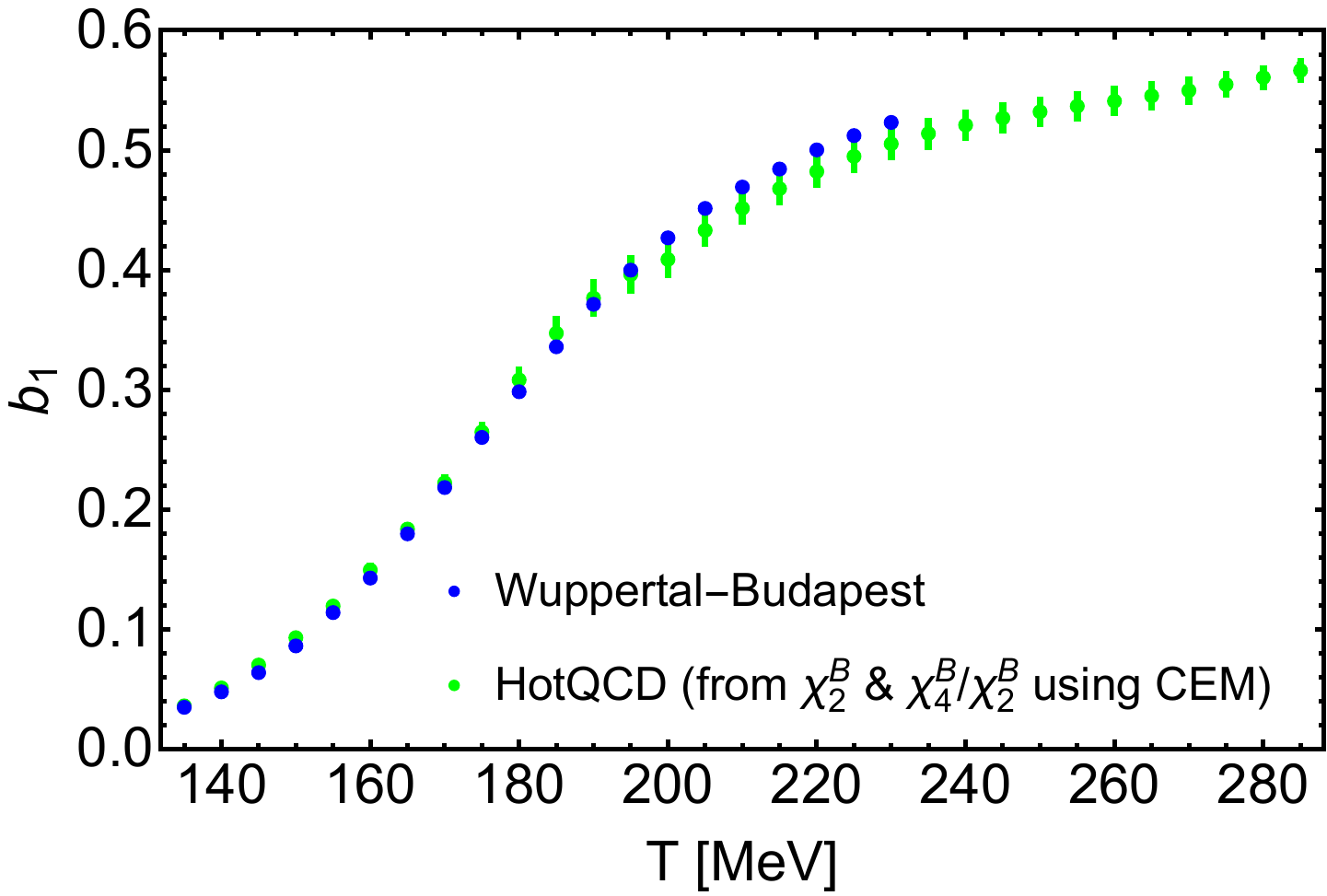}
\caption[]{(\textbf{Left):} Baryon number fluctuations $\chi^B_2, \chi^B_4, \chi^B_8$ from the lattice in comparison with the CEM 
model.  (\textbf{Right}): $b_1$ computed directly from \eq(\ref{eq:bcoeff}) by the WB collaboration, 
and~reverse engineered using CEM from
HotQCD baryon number fluctuations.
From~\cite{Vovchenko:2018zgt}.}
\label{fig:cemtest}
\end{figure}
%

An important quantity is the pseudo-critical 
temperature marking the ``phase boundary'' between the chirally broken and restored regimes.
Since the chiral transition at the physical point corresponds to an analytic crossover with a non-zero order parameter
everywhere, there are no truly distinct ``phases'' and no unambiguous definition of a transition temperature exists.
In general, definitions based on different observables will give different pseudo-critical temperatures, even in the
thermodynamic limit, contrary to the unique locations of singularities for true phase transitions. While this is 
an issue when comparing with an experimental situation, for~theoretical investigations it is convenient to stick to the 
observables representing the true order parameter in the appropriate limit, i.e.,
the susceptibility of an appropriately normalised chiral condensate in this case. 
Following as an implicitly defined function from the partition function, the~pseudo-critical temperature can be
similarly expressed as a power series in chemical potential,
\beq
 \frac{T_{pc}(\mu_B)}{T_{pc}(0)}=1-\kappa_2\left(\frac{\mu_B}{T_{pc}(0)}\right)^2+\ldots,
 \label{eq:tc_mu}
 \eeq
with $T_{pc}(0)=156.5(1.5)$ MeV~\cite{HotQCD:2018pds}.
Continuum extrapolated results for the leading coefficient are collected in Table \ref{tab:kappa2},
 the sub-leading coefficient $\kappa_4$ is  
compatible with zero at the current accuracy. This is a remarkable result telling us that up
to $\mu_B\lsi 3T$ the dependence of thermodynamic quantities on chemical potential is rather weak and can be
accurately described by a truncated leading-order Taylor series in chemical~potential.
\begin{specialtable}[H]
\setlength{\tabcolsep}{6.5mm} 
\caption{Summary of continuum-extrapolated values for $\kappa_2$ in Equation (\ref{eq:tc_mu}) .}
\label{tab:kappa2}
\begin{tabular}{ccc}
\toprule
 \boldmath{$\kappa_2$} & \textbf{Action} & \textbf{Ref.} \\ 
\midrule
0.0158(13) & \mbox{imag.}\,$\mu$, \mbox{stout-smeared staggered}& \mbox{\cite{Bellwied:2015rza}} \\
0.0135(20) & \mbox{imag.}\, $\mu$, \mbox{stout-smeared staggered}& \mbox{ \cite{Bonati:2018wdn} }\\
0.0145(25) & \mbox{Taylor, stout-smeared staggered} & \mbox{ \cite{Bonati:2018wdn,Bonati:2018nut}} \\
0.016(5) & \mbox{Taylor, HISQ} &\mbox{ \cite{HotQCD:2018pds}}\\
\bottomrule
\end{tabular}

\end{specialtable}

We now have the necessary information to obtain a conservative bound on the location of a possible critical point, which according
to \fig\ref{fig:mu_schem} sits on the pseudo-critical line of a strengthening crossover. Using the central value from
\eq(\ref{eq:tc}) for the chiral critical temperature and imposing the model-independent ordering $T_\mathrm{cep}<T_c=132$ MeV, 
the chemical potential
of a critical point must satisfy
\beq
\mu_B^\mathrm{cep} > 3.1\; T_{pc}(0) \approx 485 \;\mathrm{MeV}.
\label{eq:bound1}
\eeq

\subsection{The Search for a Critical~Point}

For any power series of a function with a given domain of analyticity in its complex argument, 
the radius of convergence gives the distance between the expansion point and the nearest singularity. 
This implies that the location $(T_c,\mu_B^c)$ of a non-analytic QCD phase transition constitutes an upper bound on the 
radius of convergence of the series \mbox{\eq(\ref{eq:press})}, or~that of any other thermodynamic function.  
Turning this around one may search for a critical point: If a finite
radius of convergence can be extracted from the pressure series for real parameter values, it should signal a phase transition.
The simplest estimator is the ratio test of consecutive coefficients, whose extrapolation 
yields the radius of convergence,
\beq
r=\lim_{n\rightarrow \infty} r_{2n}\;,\quad r_{2n}=\left|\frac{2n(2n-1)\chi^B_{2n}}{\chi^B_{2n+2}}\right|\;.
\eeq
In practice, however, only the first few coefficients are available and an extrapolation is not feasible. 
For a compilation of available results, see~\cite{Bazavov:2017dus}.
Moreover, the~ratio estimator is inappropriate for series with irregular signs and complex singularities, where it
fails to~converge.

This can be illustrated by modelling the 
lattice data in a spirit similar to the hadron resonance gas descriptions, 
such that higher coefficients become available and different scenarios can be tested for compatibility with the data.
As an example, consider the fugacity expansion of baryon number density. At~imaginary chemical potential, this 
is a Fourier series whose coefficients can be computed on the lattice without sign problem,  
\bea
\frac{n_B}{T^3}|_{\mu_B=i\theta_B T}&=&i\sum_k b_k(T) \sin(k\mu_B/T)\;,\\\
b_k(T)&=&\frac{2}{\pi}\int_0^\pi d\theta_B\; {\rm Im}\Big(\frac{n_B(T,i\theta_BT)}{T^3}\Big)\sin(k\theta_B)\;.
\label{eq:bcoeff}
\eea  
In~\cite{Vovchenko:2017gkg} a cluster expansion model (CEM) was proposed, which takes the first two coefficients as input
from a lattice calculation~\cite{Vovchenko:2017xad}, and~expresses all higher coefficients in terms of these,
\beq
b_k(T)=\alpha_k^{SB}\frac{[b_2(T)]^{k-2}}{[b_1(T)]^{k-1}}\;,\quad k=3,4,\ldots
\eeq
The $\alpha_k^{SB}$ are $T$-independent and fixed to reproduce the Stefan-Boltzmann limit.
This recursion implies that only two-body interactions are included, 
and corresponds to a truncated virial expansion which is valid for sufficiently
dilute systems. The~model now predicts all coefficients $b_{k\geq3}$ and 
allows for an all-order closed expression,
\bea
\label{eq:analyt}
\frac{n_B(T,\mu_B)}{T^3} = 
-\frac{2}{27 \pi^2} \, \frac{\hat{b}_1^2}{\hat{b}_2} \, \left\{ 4 \pi^2  \, \left[ \textrm{Li}_1\left(x_+\right) - \textrm{Li}_1\left(x_-\right) \right] + 3 \, \left[ \textrm{Li}_3\left(x_+ \right) - \textrm{Li}_3\left(x_-\right) \right] \right\}\;,\\
\mbox{with}\quad  \hat{b}_{1,2} = \displaystyle \frac{b_{1,2}(T)}{b_{1,2}^{\rm SB}}, \quad x_{\pm} = \displaystyle - \frac{\hat{b}_2}{\hat{b}_1} \; e^{\pm \mu_B/T},  \quad \textrm{Li}_s(z) = \displaystyle \sum_{k=1}^{\infty} \frac{z^k}{k^s}.\nn
\eea
All existing lattice data in the crossover regime are reproduced with excellent accuracy, 
as the examples in \fig\ref{fig:cemtest} (left) show. Moreover, \fig\ref{fig:cemtest} (right) compares the coefficient $b_1$
computed directly from its defining equation, \eq(\ref{eq:bcoeff}), with~a calculation from a combination of $\chi^B_i$, by~inverting the CEM. Note that \textit{all} orders of the fugacity expansion enter this calculation. 
The remarkable quantitative agreement 
is only possible, if~both lattice calculations (using different actions) give equivalent results and all
coefficients of the CEM are sufficiently close to the true QCD values.
\begin{figure}[H]
\hspace*{-0.6cm}
\includegraphics[width=0.39\textwidth]{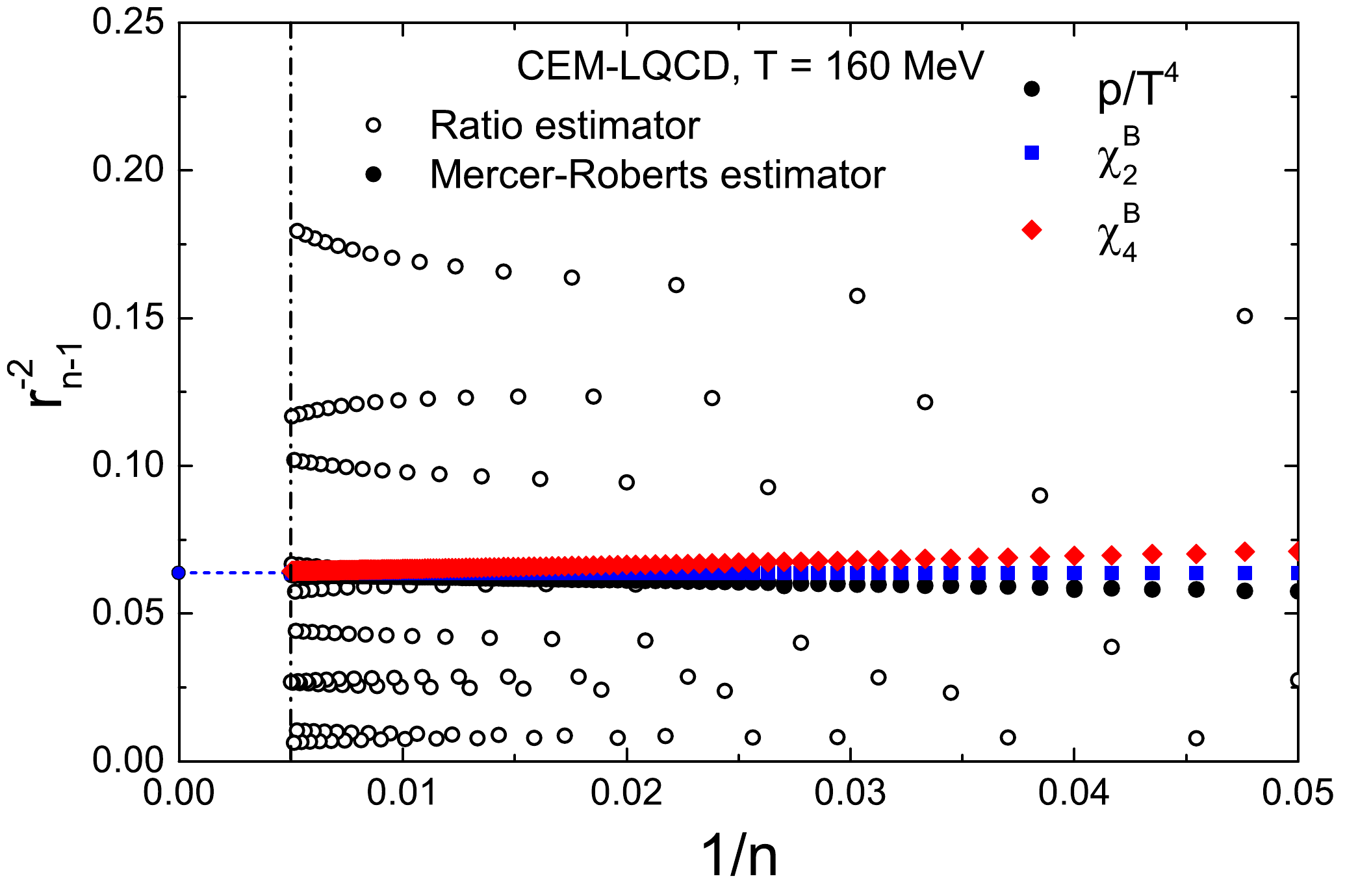}
\includegraphics[width=0.38\textwidth]{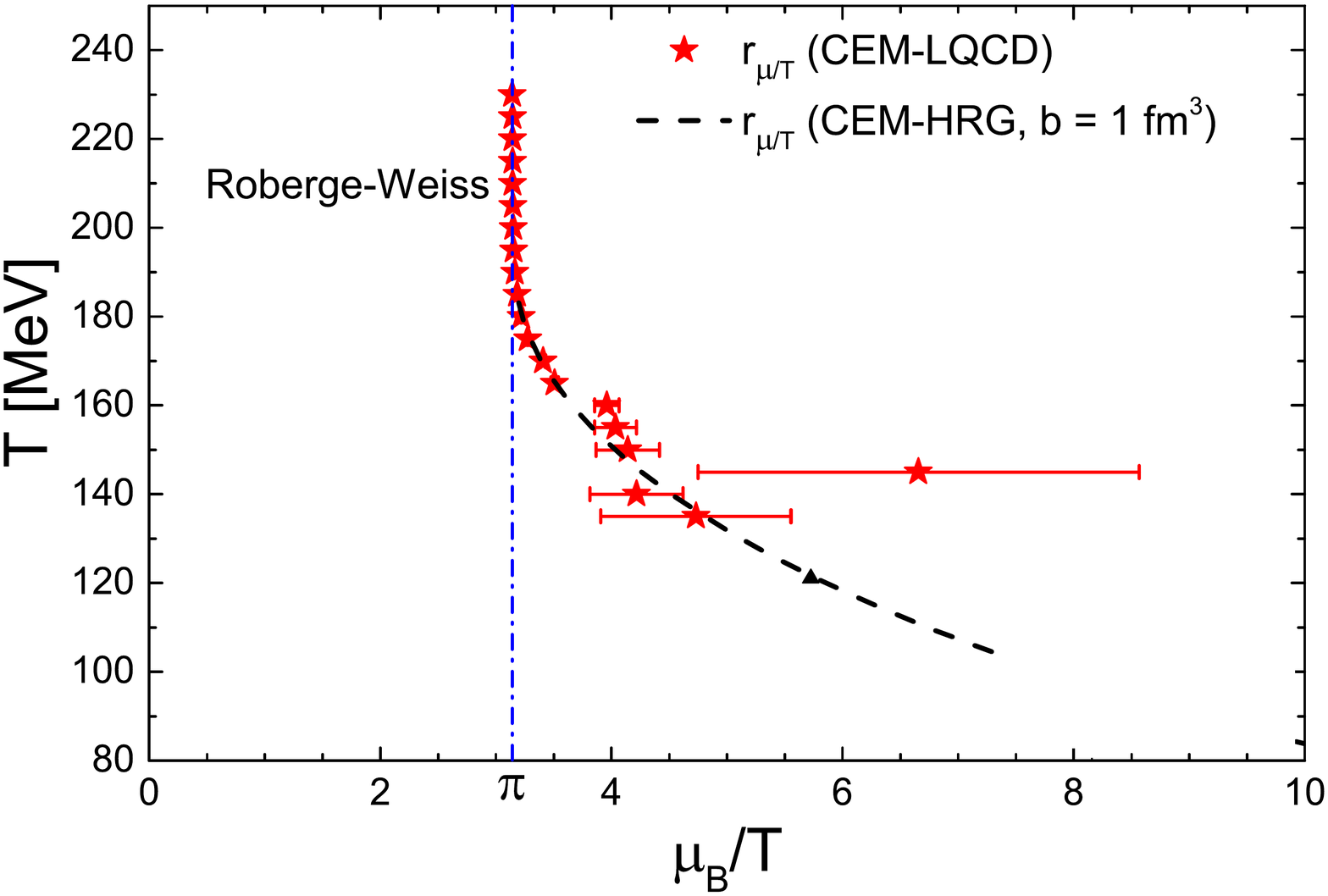}
\caption[]{(\textbf{Left}): Comparison of estimators for the radius of convergence in CEM.
(\textbf{Right}): The resulting radius of convergence as a function of $T$ predicts the RW-transition
 at imaginary $\mu_B$. From~\cite{Vovchenko:2017gkg}.}
\label{fig:cem_rad}
\end{figure}

With all coefficients of the fugacity expansion available, one can study the radius of convergence of CEM, \fig\ref{fig:cem_rad}. 
The ratio estimator fails to converge, because~of the irregular signs of higher order coefficients
(it works for equal or alternating signs). On~the other hand, the~Mercer-Roberts
estimator,
\beq
r_n=\left|\frac{c_{n+1}c_{n-1}-c_n^2}{c_{n+2}c_n-c_{n+1}^2}\right|^{1/4}\;,
\eeq
converges and extrapolates to a unique radius of convergence, even for coefficients $c_n$ pertaining to different observables,
as must be the case for a true singularity. A~number of improved estimators with faster convergence properties
has been proposed in~\cite{Giordano:2019slo}.

The red points in \fig\ref{fig:cem_rad} (right) show the resulting radius of convergence of the CEM for 
different temperatures. At~high temperatures a singularity is predicted at a distance $|\mu_B/T|\approx \pi$ and $T>T_{pc}$, 
which is quantitatively compatible with the first-order Roberge-Weiss transition in the imaginary $\mu_B$ direction.
At lower temperatures there is no Roberge-Weiss transition and the radius of convergence quickly grows. 
This is an intriguing result, given that no information from imaginary chemical potential has been used as input.
It shows that the general approach to detect a non-analytic phase transition is viable, 
provided that a good estimator for the radius of convergence 
is used and that sufficiently many 
coefficients are available (more than 100 in this case!).
The correct identification of the Roberge-Weiss transition by the radius of convergence 
conversely implies that the CEM of lattice QCD does not have another phase transition closer to the origin, i.e.,~a possible
critical endpoint in the real direction must satisfy
\beq
\mu_B^\mathrm{cep}> \pi T\;.
\eeq
\textls[-15]{This is fully consistent with \eq(\ref{eq:bound1}) but derived by completely different~methods}. 

Of course, modelling higher coefficients in terms of lower ones is not unique. 
The simplest alternative is a description of the baryon number fluctuations up to $\chi^B_8$ 
by a polynomial model, equally without singularity~\cite{Fodor:2018wul}.
Another one  is provided by a rational function model, which can account for singularities and can be applied both to 
QCD or to a chiral model with a phase transition~\cite{Almasi:2019bvl}. Equations of state 
compatible with lattice data and including an Ising critical point in predefined locations 
have also been constructed~\cite{Parotto:2018pwx}.
While the properties of models are not those of QCD, these analyses altogether do show 
that there is no sign of criticality in the real $\mu_B$ direction from the 
presently available, continuum extrapolated lattice data at zero or imaginary chemical potential, but~at
best in their higher order completions. It will thus be interesting to test higher order coefficients without modelling by
Pad\'e approximants or conformal maps as demonstrated in a Gross-Neveu model~\cite{Basar:2021hdf}, or~by 
resummation schemes in $(\mu_B/T)$ \cite{Mondal:2021jxk}. 

Having to go term by term in an expansion can be avoided, if~the radius of convergence is instead determined by 
the Lee-Yang zero~\cite{Yang:1952be,Lee:1952ig} closest to the origin. Using reweighting to real chemical potential, 
this was the strategy employed in the first lattice prediction of a 
critical point on $N_\tau=4$ lattices using unimproved rooted staggered fermions~\cite{Fodor:2001pe}. 
However, it is now understood that the closest Lee-Yang zero was caused by a spectral gap between the unrooted taste quartets
of the zeros, after~Taylor expanding the reweighting factor, rather than by a phase transition~\cite{Giordano:2019gev}. 
This is related to the general problem of staggered taste quartets splitting up when they cross a branch cut
of a rooted determinant~\cite{Golterman:2006rw}. It has then been proposed to redefine the rooted staggered determinant 
by a geometric matching procedure averaging over the quartets, after~which it can be represented as an ordinary  
polynomial~\cite{Giordano:2019gev}.  A~further development~\cite{Giordano:2020roi} 
concerns the reweighting procedure, which usually is based
on sampling with the phase quenched determinant and reweighting in the phase factor. This has a well-known
overlap problem between the sampled and reweighted ensembles. 
To get rid of this, one may neglect the imaginary part of the determinant in the partition function altogether, which 
is allowed since it will average to zero. 
One can then sample with the real part of the fermion deter\-minant and
reweight in the sign only, which has no overlap problem. Of~course, the~sign problem remains and one is still faced with
a challenging signal to noise~ratio.

Application of these new methods using the stout-smeared action on coarse $\NTau=4$ lattices appears to 
signal a Lee-Yang zero at $\mu_B\sim 2.4 T$ \cite{Giordano:2020roi}, which however is is still far from the continuum.
Simulation results on $\NTau=6$ for the renormalised chiral condensate,
\begin{equation}
  \label{eq:chiralcondensate3}
  \langle \bar{\psi}\psi\rangle_R(T,\mu) =   -\frac{m_{ud}}{f_\pi^4}\left[
    \langle \bar{\psi}\psi\rangle_{T,\mu} -\langle \bar{\psi}\psi\rangle_{0,0}
  \right]\,,
\end{equation}
are shown in \fig\ref{fig:mu_bw}, both for real and imaginary chemical potential~\cite{Borsanyi:2021hbk}. 
In the left plot, there is no steepening or narrowing of the chiral crossover yet as the real chemical potential is 
increased. In~the right plot the chemical potential is varied for fixed temperature $T=140$ MeV. 
One observes full compatibility of the 
reweighted real $\mu_B$ simulations with the analytic continuation from imaginary $\mu_B$ simulations, but~with smaller errors.
In this case, also, there is no sign of a non-analyticity, which is again consistent with the results from  
other~methods.

\begin{figure}[H]
\includegraphics[width=0.37\textwidth]{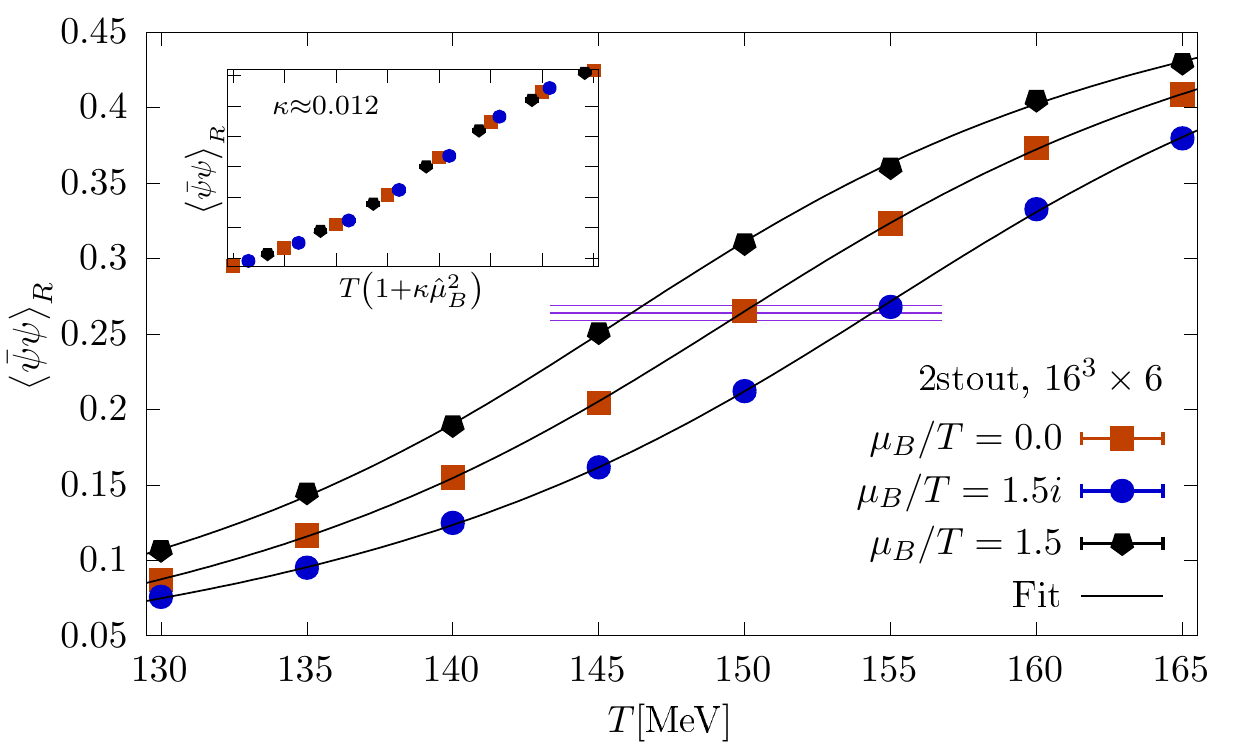}
\includegraphics[width=0.37\textwidth]{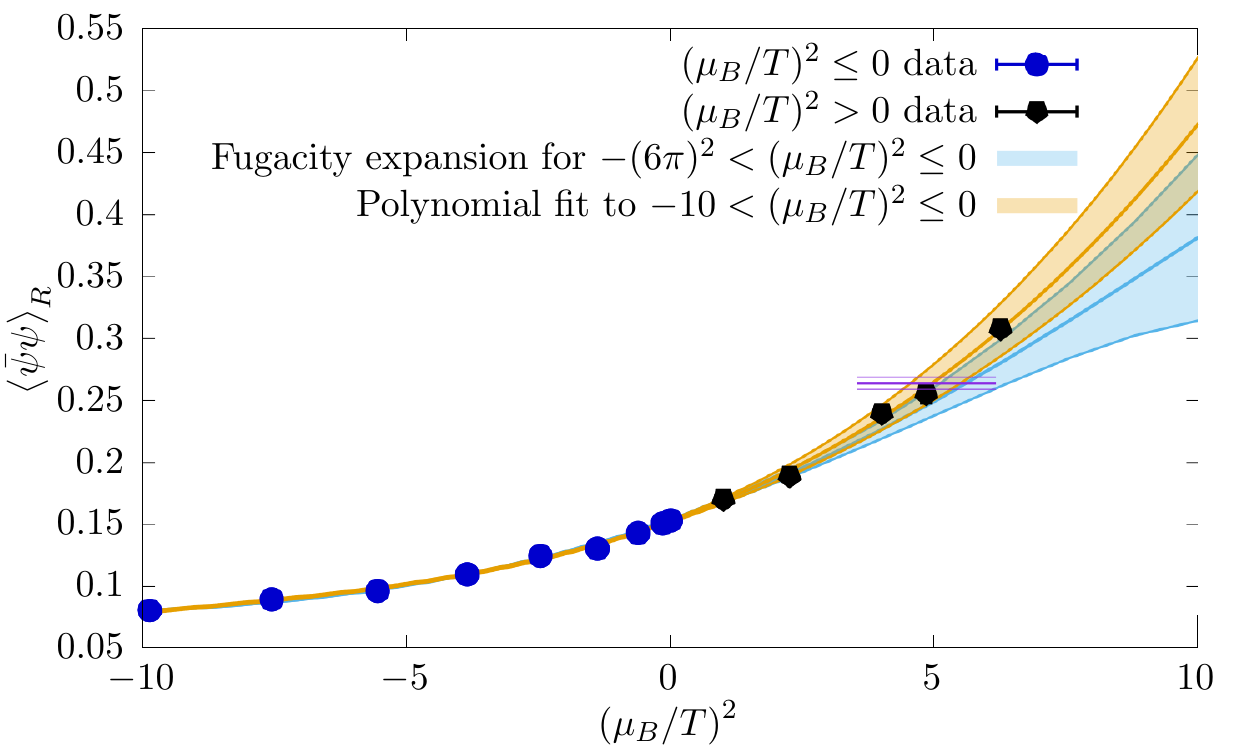}
\caption[]{Renormalised chiral condensate from simulations with stout-smeared staggered fermions 
at imaginary (blue) and real (black) chemical
potential. (\textbf{Left}): Temperature scan at various chemical potentials. 
(\textbf{Right}): $\mu_B$ scan at at $T=140$ MeV. 
Coloured bands result from analytic continuation of imaginary chemical potential data. From~\cite{Borsanyi:2021hbk}.}
\label{fig:mu_bw}
\end{figure}
\section{Conclusions}

Even though the sign problem of lattice QCD remains unsolved, there has been considerable progress over the last
few years towards phenomenologically relevant constraints on the QCD phase diagram. This is due to a number of
complementary paths of investigation, each of them refining their methods and having finer lattices as well as different
lattice actions at their~disposal. 

In the Columbia plot at $\mu_B=0$, 
the region of first-order deconfinement transitions in the heavy mass corner 
can be directly simulated, and~its $Z(2)$-critical boundary for $\Nf=2$, while not yet continuum extrapolated, 
is expected to settle in a region around $m_{PS}\sim 4$ GeV. 
The region of first-order chiral phase transitions seen in various lattice discretisations has been observed to 
shrink continuously with decreasing lattice spacing. There is numerical evidence for several $\Nf\geq 3$ 
that their $Z(2)$-critical boundary    
extrapolates to the chiral limit at finite lattice spacing, exhibiting tricritical scaling. Such first-order 
regions are not connected to the continuum and represent lattice artefacts. 
If this is confirmed by further studies, the~Columbia
plot will look like in \fig\ref{fig:continuum}.

In another recent development, the~critical temperature for the chiral transition in the massless limit has been determined
to be in the range $T\approx 126-140$ MeV, which is significantly lower than the pseudo-critical temperature at the physical
point. According to the expected relationship between the massless limit and the physical quark masses, this bounds 
a possible critical endpoint at finite baryon density to be at $\mu_B$ $\gsi$ $3 T$. An~estimate of the radius of convergence,
based on a cluster expansion model consistent with all available zero density baryon number fluctuations for physical QCD, 
is fully consistent with this.
Novel techniques to extract the leading Lee-Yang zero of the partition function, either with improved reweighting methods
or by refined series analyses, 
should be able to provide additional, independent estimates of the proximity of a phase transition to the physical point 
in the near~future.

\vspace{6pt} 



\funding{Parts of the work reported here were funded by the Deutsche Forschungsgemeinschaft (German Research Foundation),  grant number 315477589-TRR 211 (Strong-interaction matter under extreme conditions).}

\conflictsofinterest{The author declares no conflict of~interest.}

\end{paracol}
\reftitle{References}

\end{document}